\documentclass{article}

\usepackage{arxiv}

\usepackage[utf8]{inputenc} 
\usepackage[T1]{fontenc}    
\usepackage{hyperref}       
\usepackage{url}            
\usepackage{booktabs}       
\usepackage{amsfonts}       
\usepackage{nicefrac}       
\usepackage{microtype}      
\usepackage{lipsum}
\usepackage[font=small,labelfont=bf]{caption} 

\usepackage{graphicx,calc}
\usepackage{amsmath}
\usepackage[english]{babel}
\usepackage{eurosym}

\title{Buckling \textit{soft} tensegrities: fickle elasticity and configurational switching in living cells}

\author{
  M. Fraldi\\
  Department of Structures for Engineering and Architecture, University of Napoli Federico II, Italy\\
  Institute for Applied Sciences and Intelligent Systems, National Research Council of Italy\\
  \texttt{fraldi@unina.it} 
   \And
  S. Palumbo \\
  Department of Civil, Environmental and Mechanical Engineering, University of Trento, Italy\\
  \texttt{stefania.palumbo@unitn.it} \\
  \And
  A.R. Carotenuto\\
  Department of Structures for Engineering and Architecture, University of Napoli Federico II, Italy\\
  \texttt{angelorosario.carotenuto@unina.it}\\
  \And
  A. Cutolo\\
  Department of Structures for Engineering and Architecture, University of Napoli Federico II, Italy\\
  \texttt{arsenio.cutolo@unina.it}\\
  \And
  L. Deseri \\
  Department of Civil, Environmental and Mechanical Engineering, University of Trento, Italy\\
  Department of Mechanical Engineering \& Department of Civil and Environmental Engineering,\\ Carnegie Mellon University, Pittsburgh, PA, USA\\
  \texttt{luca.deseri@unitn.it}\\
  \And
  N. Pugno\\
  Department of Civil, Environmental and Mechanical Engineering, University of Trento, Italy\\
  School of Engineering and Materials Science, Queen Mary University, London, UK\\
  Laboratory of Bio-Inspired \& Graphene Nanomechanics, University of Trento, Italy\\
  Ket Lab, Edoardo Amaldi Foundation, Italian Space Agency, Rome, Italy\\
  \texttt{nicola.pugno@unitn.it}
}

\begin{document}
\maketitle

\begin{abstract}
Tensegrity structures are special architectures made by floating compressed struts kept together by a continuous system of tensioned cables. Their existence in a mechanically stable form is decided by the possibility of finding geometrical configurations such that pre-stressed tendons and bars can ensure self-equilibrium of the forces transmitted through the elastic network, the overall stiffness of which finally depends on both the rigidity of the compressed elements and the cables' pre-stress. The multiplicity of shapes that tensegrity structures can assume and their intrinsic capability to be deployable and assembled, so storing (and releasing) elastic energy, have motivated their success as paradigm --pioneeringly proposed three decades ago by the intuition of Donald E. Ingber-- to explain some underlying mechanisms regulating dynamics of living cells. The interlaced structure of the cell cytoskeleton, constituted by actin microfilaments, intermediate filaments and microtubules which continuously change their spatial organization and pre-stresses through polymerization/depolymerization processes, seems in fact to steer migration, adhesion and cell division by obeying the tensegrity construct.
Even though rough calculations lead to estimate discrepancies of less than one order of magnitude when comparing axial stiffness of actin filaments (cables) and microtubules (struts) and recent works have shown bent microtubules among stretched filaments, no one has yet tried to remove the standard hypothesis of rigid struts in tensegrity structures when used to idealize the cell cytoskeleton mechanical response. With reference to the $30$-element tensegrity cell paradigm, we thus introduce both compressibility and bendability of the struts and accordingly rewrite the theory to simultaneously take into account geometrical non-linearity (i.e. large deformations) and hyper-elasticity of both tendons and bars, so abandoning the classical linear stress-strain constitutive assumptions. By relaxing the hypothesis of rigidity of the struts, we demonstrate that some quantitative confirmations and many related extreme and somehow counter-intuitive mechanical behaviors actually exploited by cells for storing/releasing energy, resisting to applied loads and deforming by modulating their overall elasticity and shape through pre-stress changes and instability-guided configurational switching, can be all theoretically found. It is felt that the proposed new \textit{soft}-strut tensegrity model could pave the way for a wider use of engineering models in cell mechanobiology and in designing bio-inspired materials and soft robots.
\end{abstract}


\section{Introduction}
\begin{figure}[h]
\centering
\includegraphics[width=0.9\textwidth]{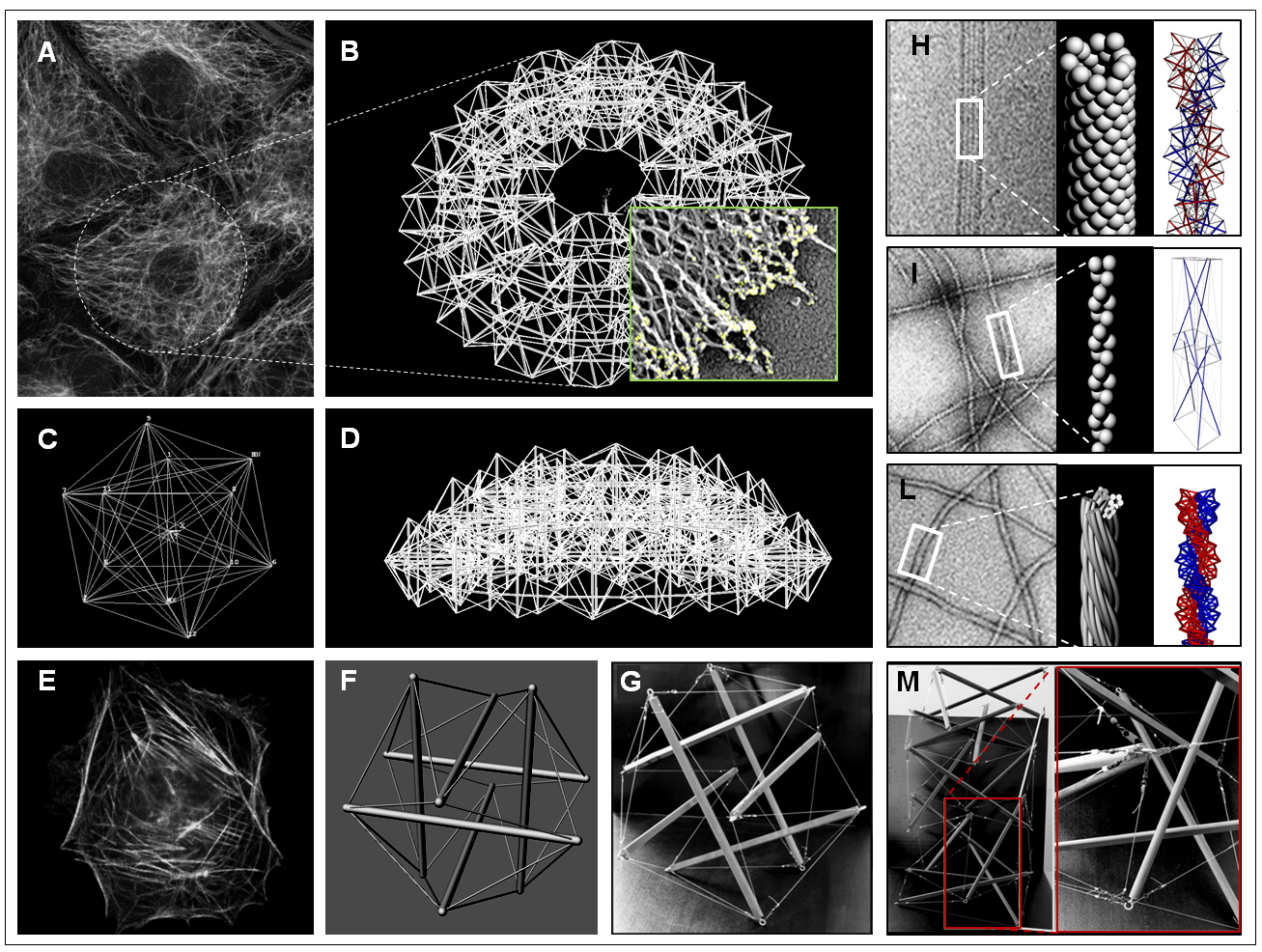}    
\caption{Synoptic panel illustrating how tensegrities can be met across the scales in cell structures. \textbf{A)} Standard microscope image of cells in which is highlighted their cytoskeleton; \textbf{B)} possible tensegrity-based multi-modular model of a cell unit capable to replicate the complex geometry of the filaments network (green window) and lateral view of the model (\textbf{D}); \textbf{C)} top view of a single cell Finite Element model made of an enriched icosahedral tensegrity whose nodes are all interlaced by filaments to capture details of the cytoskeleton architecture (\textbf{E)}); \textbf{F)} $30$-element tensegrity structure utilized in the present work to idealize the cell mechanical behavior; \textbf{G)} macroscopic handmade toy system of a $30$-element tensegrity structure;  \textbf{H-L)} microtubules, actin and bundles of microfilaments whose structures can be modeled by means of piled tensegrity modules (shown on the right) such as the so-called \textit{Snelson tower} (\textbf{M)} reports one constructed by the authors) which is built up by repeating --and properly connecting-- $30$-element tensegrity units along a prescribed direction.}
\label{fig.synoptic}
\end{figure}

The single cell can be thought as a unitary element embedded in a complex entangled space, able to continuously receive and respond to external biochemical and mechanical signals \cite{mechbook}. Recently, wide interest has been addressed to the role played by the cell biomechanics and mechanobiology in the mechanotransduction processes \cite{C3SM52769G} that seem to regulate many important cellular functions --such as proliferation, differentiation, migration as well as neoplastic mutations-- by means of viscoelastic properties \cite{Fraldi_2015,Fraldi_2017} and mechanical-driven cells morphological changes\cite{Stamenovic,nih}.\\
From the mechanical point of view, cells can be modeled as continuum media when the smallest length scale of interest is significantly larger than the dimensions of their microstructure. In such a case, averaged constitutive laws are applicable to the whole cell or cellular compartments at the macroscopic level and the predictions furnished by these kind of models strictly depend on the suitability of the chosen constitutive laws, the effects of the underlying microstructure resulting averaged and so necessarily at least in part lost. While continuum approaches can be helpfully adopted at meso--/macro--scopic scales, they appear hence less useful when one aims to investigate the way in which stresses and strains induced on the cell are transmitted through the discrete subcellular components or if the interest is to understand how internal mechanical forces govern the cell behavior by modulating the pre-stress level in the cytoskeletal fibers in turn influencing the overall cell actual stiffness, the stored (internal) energy and the adhesion and migration cell mechanisms. For these reasons, it is commonly accepted that continuum mechanics-based models which conceive the cellular apparatus as a force-bearing cortical membrane including a viscoelastic cytoplasm, by ignoring its inner microstructure, lack the ability to catch distributions and channeling of forces within the cell, that instead seem to importantly account for cell structural stability \cite{mofrad2006cytoskeletal} as well as for characteristic phenomena such as the so-called "actions at a distance" \cite{Ingber_2003, Wang_2009}. Several experimental evidences have shown that the transmission of mechanical forces is mainly borne by the cytoskeleton architecture which \textit{de facto} seems to behave like a discrete mechanical network \cite{Wang2001MT,Ingber_2003}, that reacts to the mechanical stimuli coming from the surrounding environment --e.g. cell-cell and cells-ECM (extra-cellular matrix) interactions-- through a global reorientation and rearrangement of its elements \cite{Tee_2015,Xu_2016,Xu_2018}. The crossroad in interpreting the structural principles underlying the cytoskeleton behavior has been offered by Donald E. Ingber's intuition, according to which cells might obey tensegrity structural principles \cite{Ingber_1981,Ingber_1985,Ingber_1993b}. Many works have demonstrated that tensegrity structures can be also traced at different scale levels in biological matter, organized hierarchically with nested or self-similar architectures. They can be in fact recognized in the inner structure of actin microfilaments, in the sub-regions of cells, at overall cell cytoskeleton level (see Figure \ref{fig.synoptic}) and observed across the scales up to the macroscopic level of tissues such as tendons and muscles, as well as in the skeletal systems of vertebrates to preside over locomotion and load-bearing functions \cite{Gray}.

However, differently from other structural or geometrical paradigms at the basis of hierarchical constructs \cite{FRALDI20131310}, tensegrity systems exhibit additional intrinsic features which biological materials seem to very helpfully exploit to maximize selected performances and optimize mechanotrasduction signaling patterns. For example, it can be seen that, by scaling in size a tensegrity unit or by tuning the pre-stress of its elements, the resulting mechanical response may nonlinearly vary to span across orders of magnitudes in terms of elastic moduli. This opens the way to a wide range of different possible cell mechanical behaviors, that are also related to stored energy releasing, visco-elastic softening of the constituents and polymerization/depolymerization-mediated phenomena. The concurrence of these processes could result in snap-through and post-buckling effects involving cytoskeletal actin filaments, microfilaments and microtubules, as we will show below. In this manner, migration, stiffness tuning, duplication and adhesion could be somehow all regarded --and hence modeled-- as peculiar cell abilities allowed by mechanisms intrinsically exhibited by tensegrity systems.\\
Tensegrity is defined as a system in a stable self-equilibrated state comprising a discontinuous set of compressed components inside a continuum of tensed ones, so that all the structural members already experience an internal state of stress --i.e. a state of pre-stress-- before the application of any external force \cite{motro, skelton}.
From the mechanical point of view, the so-called form-finding problem for elastic tensegrity structures --aimed to find geometrical configurations in which self-equilibrated stress states are associated to compatible deformations of cables and struts-- is a fully nonlinear problem, involving large displacements, finite strains and hyperelastic constitutive laws. Large deformation regimes are always required for essentially two reasons: the first is that one has to explore families of geometrical shapes significantly far from the one selected as the reference configuration, as an effect of external loading, and the second is that the working principle on which the tensegrity concept is based needs the presence of pre-stretched (or equivalently elastically pre-stressed) elements. In the vast majority of the examples found in literature to model macroscopic systems (see, for example, $3$-strut tensegrity structures in large deformation in \cite{Fraternali2015} and tensegrity-based robot prototypes \cite{Liu_2017} as well as in the cases where these structures are used to describe mechanics of single cells \cite{Stamenovic1996,Stamenovic} or of their constituents \cite{Luo2008}), the elastic constitutive laws are however treated as an ancillary issue, due to the fact that the cables are assumed to deform by experiencing tensile regimes and the compressed struts are generally considered as rigid components. At the best authors' knowledge, except in \cite{violinbow}, in the rare literature cases where struts deform, they are modeled by implicitly adopting de Saint Venant-Kirchhoff constitutive behavior \cite{Fraternali2015} or as perfectly rigid up to a critical compressive load, then buckling and thus involving a finite flexural rigidity \cite{Coughlin1997,Coughlin1998,Stamenovic2000,Volokh2000} . Actually, the cytoskeleton can be interpreted as a tensegrity-like network comprising three main polymeric components: actin microfilaments, microtubules and intermediate filaments. Tensile pre-stress in actin microfilaments is partially actively generated through ATP-driven processes (contractile actomyosin motors) and partially passively generated through cell spreading on ECM and cytoplasmic swelling pressure. It is balanced both by compressive-bearing microtubules and by the traction forces exerted on the cell by surrounding cells and ECM, through specific integrin sites (focal adhesions), with variable contributions depending on the degree of cell adhesion \cite{Stamenovic_2001,nih}.   
Several works have demonstrated that a $30$-element cable-and-strut tensegrity structure can be assumed as a good candidate for reproducing the cytoskeletal apparatus, able to account, at least qualitatively, for a number of properties exhibited by cells \cite{Stamenovic1996,Coughlin1997,Coughlin1998,Stamenovic2000,Volokh2000}. In static conditions, this architecture has been studied by modeling actin microfilaments as linearly elastic (tensed) cables and microtubules as rigid \cite{Stamenovic1996} or as elastic slender struts able to buckle under compression \cite{Coughlin1997,Coughlin1998,Volokh2000}.
However, experimental studies have shown the possibility that single actin microfilaments and, mainly, higher-order structures deriving from their assembling in bundles --namely stress fibers-- could exhibit nonlinear behaviors \cite{Liu2002,Deguchi2006}, and that intermediate filaments, not included in the aforementioned works, could play a significant role at high levels of strain and may also represent a lateral tensile support for microtubules, enhancing their capability to resist buckling under compression \cite{Brodland1990}.
As a matter of fact, by estimating the actual ratio between axial stiffness of filaments and microtubules in living cells, one finds that it can approach values which tend to the unity or differ from it for less than an order of magnitude. As a consequence, especially when the cytoskeleton is extremely stretched --for instance during cell adhesion-- buckling coupled with high contractions of microtubules might take place, this forcing us to abandon linear elasticity in the form of de Saint Venant-Kirchhoff law, that produces inconsistent results at high compressive stretches \cite{holzapfel}.
Motivated by these observations and encouraged by the evidence according to which the cytoskeletal network undergoes nonlinear deformations and large displacements in the most of the cells physiological processes \cite{Gardel2008}, such as spreading, adhesion and isolated or collective migration, in the present work we re-examine the $30$-element tensegrity paradigm by providing both geometrical and constitutive nonlinearities. In particular, to overcome the limitations above mentioned and build up a flexible strategy for more faithfully predicting some experimentally observed cell cytoskeleton mechanical behaviors, the effects of three different choices of nonlinear elastic models on the response of a  \textit{soft}-strut tensegrity system are firstly investigated, within a general theoretical framework. This allowed to prove that: \textit{i}) the standard de Saint Venant-Kirchhoff law is mechanically incompatible as the struts feel high axial contractions, \textit{ii}) the Hencky model, naturally involving the true (logarithmic) strain implemented in most of the commercial Finite Element codes when enabling large deformations, is consistent for struts and, providing elastic softening at prescribed pre-stretch levels, could produce/anticipate switch of the whole structure from stable to unstable configurations, accompanied by loss of symmetry, and \textit{iii}) the results obtained by using classical neo-Hookean and Hencky behaviors can highlight significant discrepancies in terms of both form-finding and overall stiffness, even though the reference tensegrity geometry is initially the same.
Successively, by considering a cell cytoskeleton as a $30$-element tensegrity module with actual deformability of both microtubules (struts) and actin microfilaments (cables), the overall response of the system, in terms of generalized stress-strain relations and associated varying stiffness, is obtained analytically for three relevant load cases, that is elongation/contraction, shear and torsion, under different pre-stress conditions. Finally, by implementing the model in a Finite Element code, the same above mentioned boundary conditions and loads are prescribed to the system by additionally activating the possibility to combine axial deformations of cables and struts with buckling of the compressed elements, so determining a variety of further complex responses characterized by instability, softening and loss of shape symmetry which allow to quantitatively predict stiffness measurements found through \textit{in vitro} experimental tests and resemble behaviors actually observed in severely stretched cells \cite{Coughlin1997,Coughlin1998,Volokh2000}.
\section{Cell cytoskeleton modeled as 30-element \textit{soft}-strut tensegrity}\label{lb1}

The cell cytoskeleton is a deforming, moving and self-assembling architecture which plays a key role in essentially any cellular biological process, by providing structural stability, determining the cell shape and constituting the network filtering most of the relevant mechanotransduction signals which decide on cell migration, adhesion and division.
As recalled above, among several models proposed in the literature, the $30$-element tensegrity represents the simplest and most effective microstructural paradigm to describe the cell biomechanical behavior. The idea is to view the cytoskeleton as an interconnected system of actin microfilaments and microtubules which distributes forces within the cell, by dynamically balancing compression and tension of its constituting pre-stressed elements in suspended and adherent configurations, as well as during cell locomotion. Accordingly to the tensegrity principle, the internal forces (pre-stress) confer to the cell the needed shape stability and stiffness to continuously adapt overall cell elastic properties and cytoskeleton architecture to respond to biomechanical stimuli, allow adhesion and facilitate spreading.
More specifically, by starting from the idea by Ingber, the cell's cytoskeleton is here regarded as a $30$-element tensegrity system --with regular icosahedral geometry--  which comprises $6$ discontinuous (not directly in contact) pre-compressed struts, representing the cytoskeletal microtubules, whose ends are interconnected through $24$ pre-tensed cables, corresponding to the actin microfilaments (see Figure \ref{fig.synoptic}-\textbf{F}). Standard hypotheses of torqueless and frictionless ball-joints were assumed, the stable tensegrity configuration in the absence of external forces being found in correspondence of a set of tensed members and compressed elements in self-equilibrium. To take into account the actual axial and bending deformability of each element of the cell cytoskeleton, we therefore update standard previous models, \textit{ad hoc} conceiving a new \textit{soft}-strut $30$-element tensegrity structure which includes both large deformations and nonlinear elastic behavior of the constituents, thus \textit{ab imis} accordingly rewriting the form-finding problem and analyzing the response of the system under elongation/contraction, shear and torsion conditions.
\begin{figure}[h]
\centering
\includegraphics[width=0.99\textwidth]{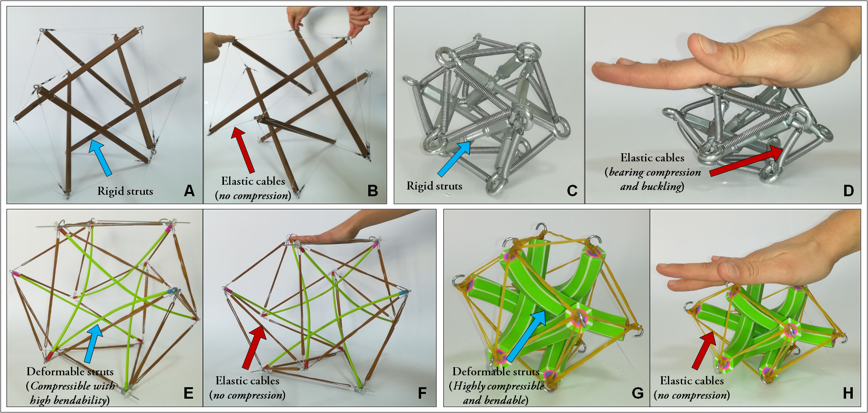}
\caption{Four handmade $30$-element tensegrity toy systems in their natural (self-equilibrated and pre-stressed) reference and deformed (slightly crushed) configurations, built up by using materials and elements such as to replicate all the relevant modulations of axial stiffness of cables and bars and bendability of struts. \textbf{A)} and \textbf{B)}: reference and deformed configurations of a standard $30$-element tensegrity, with rigid bars and tensioned cables unable to bear compression; \textbf{C)} and \textbf{D)}: reference and deformed configurations of a quasi-classical $30$-element tensegrity, with rigid steel struts and tensed metallic elastic springs, capable to support compression and to undergo buckling; \textbf{E)} and \textbf{F)}: reference and deformed configurations of a $30$-element tensegrity with axially rigid but bendable struts and tensed rubber elastic cables; \textbf{G)} and \textbf{H)}: reference and deformed configurations of a \textit{soft}-strut $30$-element tensegrity made of axially deformable and bendable (rubber) bars and tensed elastic cables.}
\label{fig.4TSs}
\end{figure}
From the mechanical point of view, the competition among axial deformability of both cables and struts and bendabilty of the same struts can produce very different results in soft-tensegrity structures undergoing growing pre-stress levels or reacting to applied loads of increasing magnitude. This is coherent with what observed in human cells, where the cytoskeleton is forced to continuously change its architecture and --with it-- the effective ratios between axial stiffness and bending rigidity of its elements, as a consequence of polymerization/depolymerization processes and of the interactions of the protein filaments, embedded in the cytosol, with the ECM. As a result, these events can in fact make tip the scales in favor of structural configurations alternatively more prone to make prevalent the axial deformability than the bendability of the struts and \textit{vice versa}.\\
Motivated by these considerations, we propose to analyze the form-finding and the behavior of the soft-tensegrity systems under selected loading conditions by separating their response in cases dominated by the axial deformation of cables and struts (high bars' bending stiffness), those governed by highly bendable struts and the intermediate situations, where axial and flexural stiffness of the cytoskeleton elements compete (see Figure \ref{fig.4TSs}).
This allowed us to recognize two sole mechanically relevant classes of tensegrity deformations and associated equilibria: the symmetry-preserving one, where both struts and cables can axially deform --also significantly-- without violating the expected symmetries imposed by the initial geometry, the pre-stress and the boundary conditions, the polyhedral regular shape of the tensegrity being kept preserved in absence of external loads, and the case of loss of (local and/or global) symmetry, where buckling instability combined with axial deformability of struts and/or change of the overall shape of the tensegrity, associated to configurational switching, can all take place producing somehow symmetry losses, both when deforming under applied loads and at increasing pre-stress in self-equilibrium states.

\subsection{Brief remarks on the kinematics at finite strain}
With the aim of dealing with finite deformations of struts and cables, we make reference to the general form of the strain measure given by the Seth-Hill formula \cite{Hill_1968,Bigoni2012}:
\begin{equation}\label{SethHill}
\textbf{E}_m := \begin{cases}
\dfrac{1}{m}\left( \textbf{U}^{m} - \textbf{I}\right) \quad m \neq 0 \\
\\
\ln  \textbf{U} \quad \quad  m = 0
\end{cases}
\end{equation}
where $ \textbf{I} $ represents the second-order identity tensor and $\textbf{U}$ is the right (or material) stretch tensor, resulting from the polar decomposition of the deformation gradient such as $ \textbf{F} = \textbf{R} \textbf{U}$, $ \textbf{R} $ being the rigid rotation. In particular, one of the strain measures preferred in the present work is the (Lagrangean) Hencky's one --also known as \textit{true strain}-- corresponding to the limit case of $ m\to 0$ shown in \eqref{SethHill}:
\begin{equation}
\label{Hencky tensor}
\textbf{H} := \textbf{E}_0 = \lim_{m \rightarrow 0} \textbf{E}_m = \lim_{m \rightarrow 0} \dfrac{1}{m}\left( \textbf{U}^{m} - \textbf{I}\right) = \ln  \textbf{U}, 
\end{equation} 
By referring to the one-dimensional case in which the tensegrity elements actually will be found, the compatibility equation (\ref{Hencky tensor}) gives: 
\begin{equation}
\label{hencky}
H = \ln {\lambda},
\end{equation}
this result being also derivable, in 1D, by direct integration of the incremental strain $dH = dl/l $ between the extremes $l_r$ and $l_c$, respectively the reference and the current length of the element. Herein, $\lambda := l_c/l_r = 1 + \varepsilon_{eng} $ is indeed the stretch and $ \varepsilon_{eng}:= \left( l_c - l_r\right) /l_r $ represents the classical engineering strain. The adoption of the logarithmic strain is in particular motivated by the possibility of conveniently including pre-strains in an additive manner, by exploiting the well-known properties of the logarithms. In fact, in presence of a harboring stretch (whose nature can be either elastic or inelastic), the usual nonlinear multiplicative decomposition of stretches can be written by means of an additive linear superposition. For instance, named $L$, $L^*$ and $l$ the lengths assumed by a one-dimensional right element respectively in a reference, a pre-stretched (intermediate) and a current configuration, the total logarithmic strain results:
\begin{equation}
\label{additività}
H = \ln \frac{l}{L} = \ln \frac{L^*}{L} \frac{l}{L^*} = \ln \frac{L^*}{L} + \ln \frac{l}{L^*} = H^* + \Delta H,
\end{equation}
where $ H^* := \ln\left( L^*/L\right)  $ is the pre-strain contribution, while $ \Delta H := \ln\left( l/L^*\right)  $ is the additional increase of deformation.
It is worth to highlight that the convenient choice of the logarithm measure of the strain does not preclude to go back anytime to other arbitrary strain measures belonging to the Seth-Hill class, as well as any classical relations --for instance a hyperelastic law written as a function of the stretch $\lambda$-- can be equivalently rewritten in terms of true strain $H$ by simply recalling that $\lambda=e^{H}$.

\subsection{Deformable struts and cables: the need to abandon linear elasticity}\label{lb2}
To describe the kinematics of both cables under tensile loads and soft (axially compressible and bendable) struts of the 30-element tensegrity structure used to model the cell cytoskeleton, we adopt stretch and the associated Hencky's logarithmic strain measure. Hyperelastic behavior and large deformations are thus assumed for all the elements of the structural cell network, for the first time including concurrent buckling and contraction of struts and so more faithfully taking into account the actual axial stiffness ratio of compressed microtubules and tensed actin filaments and microfilaments. However, the possibility of struts to be axially deformed forces us to abandon the approach commonly used in the literature for analyzing the mechanical behavior of tensegrity systems \cite{skelton,Fraternali2015} --e.g. to consider finite strain but \textit{linear} de Saint Venant-Kirchhoff laws for both cables and bars-- the linear stress-stretch relations leading in fact to physically incompatible results as moderately large contraction levels are attained. We therefore rewrite the equations governing the mechanics of \textit{soft}-strut tensegrities, by examining two constitutive hyperelastic laws for tendons and bars, first introducing both Hencky-type and neo-Hookean isotropic strain energy functions and then deriving the uniaxial stress-stretch equations for both the cases in terms of first Piola-Kirchhoff stress versus stretch (or associated logarithmic strain). To highlight the need to abandon the above mentioned classical approach used for tensegrity structures that assumed large strain and linear constitutive laws, below we also briefly recall the de Saint Venant-Kirchhoff model to show the inconsistency of it for \text{soft} struts, both in the standard cases of $\textbf{E}_1$ and $\textbf{E}_2$ chosen as alternative measures of strains.

\begin{figure}[h]
\centering
\includegraphics[width=0.7\textwidth]{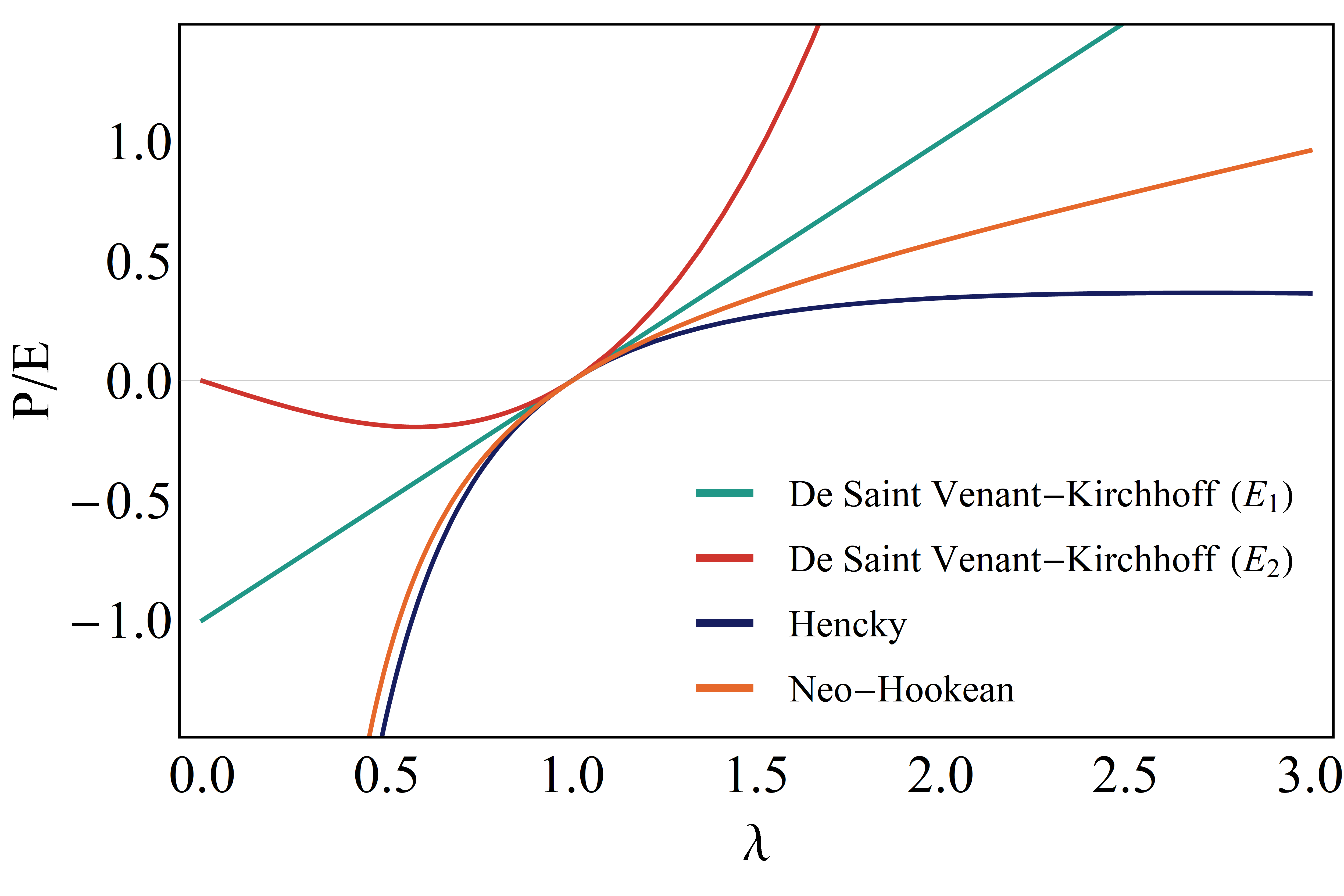}
\caption{Longitudinal nominal stress $P$ --normalized with respect to the Young modulus $E$-- vs longitudinal stretch $\lambda$ in case of Hencky, neo-Hookean (with Poisson's ratio $\nu\rightarrow0.5$) and de Saint Venant-Kirchhoff models, in one-dimensional stress regime. Note the paradoxical situations reproduced by de Saint Venant-Kirchhoff models as the contraction (in a strut) increases, for both the cases in which $\textbf{E}_1$ (the engineering strain) and $\textbf{E}_2$ (Green-Lagrange strain) are considered, respectively giving that, as $\lambda\rightarrow0$, the stress approaches a (negative) finite value and zero.}
\label{fig.PL}
\end{figure}

\subsubsection{Inadmissible linear elasticity for highly deformable struts: the inconsistency of de Saint Venant-Kirchhoff models}\label{lblb3}
The de Saint Venant-Kirchhoff (SVK) model represents the simplest three-dimensional continuous generalization of the linear (Hookean) elastic spring at large deformations. By considering the generic strain measure \eqref{SethHill}, the SVK strain energy density can be written as:
\begin{equation}\label{DSVK}
\Psi_{SVK}\left(\textbf{E}_m\right)=\frac{1}{2}\,\textbf{E}_m:\mathbb{C}:\textbf{E}_m=\frac{E}{2(1+\nu)}\,\left[ I_1\left(\textbf{E}_m^2\right)+\frac{\nu}{1-2\nu}\, I_1\left(\textbf{E}_m\right)^2\right].
\end{equation}
where $E$ is the Young modulus and $\nu$ the Poisson's ratio. With respect to the uni-axial stress regime which struts and cables undergo, by selecting $\textbf{E}_1=\textbf{U}-\textbf{I}$ from the Seth-Hill formula, the deformation in the generic cylindrical element of the tensegrity in a Cartesian frame takes the form $\textbf{E}_1=Diag\left\{\lambda_R-1,\lambda_R-1,\lambda_L-1\right\}$, so that the expression of the nominal (first Piola-Kirchhoff) stress components read:
\begin{align}
P_R&=\frac{\partial\Psi_{SVK}}{\partial\lambda_R}=\frac{E}{(1+\nu)}\left[\lambda_R-1+\frac{\nu}{1-2\nu}(2\lambda_R+\lambda_L-3)\right],\label{PDSVKR}\\
P_L&=\frac{\partial\Psi_{SVK}}{\partial\lambda_L}=\frac{E}{(1+\nu)}\left[\lambda_L-1+\frac{\nu}{1-2\nu}(2\lambda_R+\lambda_L-3)\right].\label{PDSVKL}
\end{align}
in which $L$ and $R$ denote the longitudinal and the transverse (radial) directions, respectively. To have uni-axial stress, say along the element axis, the transverse nominal stress must be vanishing, i.e. $P_R=0$. As a consequence, the following relation between transverse and longitudinal stretches has to be imposed:
\begin{equation}\label{lDVSK}
P_R=0 \Rightarrow \lambda_R=1+\nu-\nu\lambda_L
\end{equation}
which, inserted into the equations \eqref{DSVK} and \eqref{PDSVKL}, allows to obtain the following expressions for the strain energy function and the nominal stress, that highlight the direct analogy with the case of small strain and linear elasticity:
\begin{equation}\label{linear}
\Psi_{SVK}=\frac{1}{2}E\left(\lambda_L-1\right)^2,\quad P_L=E\left(\lambda_L-1\right).
\end{equation}
On the other hand, when $\textbf{E}_2=\left( \textbf{U}^2-\textbf{I}\right)/2$ is adopted as alternative measure of strain \cite{holzapfel}, one obtains $\textbf{E}_2=Diag\left\{\lambda_R^2-1,\lambda_R^2-1,\lambda_L^2-1\right\}$, whose associated stresses are:
\begin{align}
P_R&=\frac{\partial\Psi_{SVK}}{\partial\lambda_R}=\frac{E}{2(1+\nu)}\lambda_R\left[\lambda_R^2-1+\frac{\nu}{1-2\nu}(2\lambda_R^2+\lambda_L^2-3)\right],\label{PDSVKRE2}\\
P_L&=\frac{\partial\Psi_{SVK}}{\partial\lambda_L}=\frac{E}{2(1+\nu)}\lambda_L\left[\lambda_L^2-1+\frac{\nu}{1-2\nu}(2\lambda_R^2+\lambda_L^2-3)\right].\label{PDSVKLE2}
\end{align}
As above, to guarantee the uni-axial longitudinal stress regime, it follows that:
\begin{equation}\label{lDVSKE2}
P_R=0 \Rightarrow \lambda_R=\sqrt{1+\nu-\nu\lambda_L^2}
\end{equation}
whose substitution into the equation \eqref{PDSVKLE2} provides: 
\begin{equation}\label{PlE2}
P_L=\dfrac{E}{2}\lambda_L\left(\lambda_L^2-1\right).
\end{equation}
Although the two versions of the de Saint Venant-Kirchhoff model obtained obove converge to the linear case as the stretch tends to one (limit of small strains), both the equations for the stress \eqref{linear} and \eqref{PlE2} fail from moderately high up to high contraction levels, producing paradoxical results as $\lambda_L\to0$ (see Figure \ref{fig.PL}), in a case giving finite --rather than infinite-- values of compressive stress for vanishing $\lambda_L$ and in the case of \eqref{PlE2} exhibiting a non-monotonic stress-stretch curve in the interval $0<\lambda_L<1$, the stress relaxing starting from the critical value $\lambda_L=\sqrt{1/3}$, finally approaching to zero as $\lambda_L\rightarrow0$ \cite{holzapfel}. Therefore, although the de Saint Venant-Kirchhoff model has been extensively adopted up to now in the literature to treat the mechanics of tensegrity structures \cite{skelton,Fraternali2015}, the hypothesis of deformable (contracting and bending) struts obliges to renounce to linear elastic laws (SVK models) to avoid inconsistent results. In what follows, all the analyses of soft-strut tensegrity systems will be hence performed by making reference to consistent hyperelastic laws and in particular to Hencky and neo-Hookean materials.

\subsubsection{Uni-axial stress in Hencky and neo-Hookean hyperealstic cables and struts}\label{lblb1}

If a generic isotropic, homogeneous and nonlinearly elastic element of a tensegrity structure is compressed (strut) or tensed (cable), the deformation gradient $\textbf{F}$ can be assumed in diagonal form as $\textbf{F} = Diag\left\lbrace \lambda_R, \lambda_R, \lambda_L \right\rbrace$, $\lambda_L:=l/L$ being the longitudinal stretch and $\lambda_R:=r/R$ the transverse (radial) one, capital and lowercase letters denoting reference (stress-free) and current (deformed) configurations, respectively. Accordingly, the Hencky's strain tensor can be expressed as $ \textbf{H} = Diag \left\lbrace H_R, H_R, H_L \right\rbrace$, with $H_R := \log \lambda_R$ and $H_L := \log \lambda_L$. The hypothesis of uni-axial stress regime implies that the sole nonzero component of the Cauchy (true) stress tensor $ \boldsymbol{\sigma} $ --and consequently of the corresponding Kirchhoff and first Piola-Kirchhoff (or nominal) stress tensors $\boldsymbol{\tau}$ and $\boldsymbol{P}$-- is the longitudinal one; more precisely one has:
\begin{equation}
\label{stresses}
\sigma_L = \dfrac{F}{a}, \quad
P_L = \dfrac{F}{A} = \lambda_R^{2} \sigma_L, \quad \tau_L = J \sigma_L, 
\end{equation} 
where $a$ and $A$ are respectively the current and the nominal cross-sectional areas, related each other through the radial stretch $ \lambda_R $, while $ J = \det\,\textbf{F}=\lambda_R^2 \lambda_L $, so that the general relations $\boldsymbol{\tau} = J \boldsymbol{\sigma} $ and $ \textbf{P} = \boldsymbol{\tau}\textbf{F}^{-T}$ hold true. As well-known, the strain tensor $ \textbf{H} $ and the stress tensor $ \boldsymbol{\tau} $ are work-conjugate \cite{Hill_1968,Plesek2006}, so that the Hencky's strain energy function can be introduced in the form \cite{Xiao}:
\begin{equation}
\label{Hencky SED}
\Psi_H \left( \textbf{H}\right)=\frac{1}{2}\,\textbf{H}:\mathbb{C}:\textbf{H}=\mu\, I_1\left(\textbf{H}^2\right)+\frac{\Lambda}{2}\, I_1\left(\textbf{H}\right)^2,
\end{equation}
where $ \mathbb{C} $ is the fourth-order tensor of the tangent elastic moduli. By considering an isotropic material, it is given by $ \mathbb{C} = 2 \mu \mathbb{I} + \Lambda \textbf{I} \otimes \textbf{I}$, $ \mu $ and $ \Lambda $ denoting the first and the second Lamè constants, respectively, while $\mathbb{I}$ is the fourth-order identity tensor and $ I_1 $ indicates the first invariant for any generic tensor $ \textbf{A} $, i.e. $ I_1(\textbf{A}) = tr (\textbf{A}) $. Then, the Kirchhoff (logarithmic) stress-strain linear law is readily obtained as \cite{Anand1979,Plesek2006}:
\begin{equation}
\label{Constitutive law}
\boldsymbol{\tau}=\frac{\partial \Psi_H \left( \textbf{H}\right)}{\partial \textbf{H}}= \mathbb{C} : \textbf{H}
\end{equation}
or, in terms of components:
\begin{align}
\label{tauR}
&\tau_R = 2 \mu H_R + \Lambda \left( 2 H_R + H_L \right) = 2 \mu \log \lambda_R + \Lambda \log J, \\
\label{tauL}
&\tau_L = 2 \mu H_L + \Lambda \left( 2 H_R + H_L \right)  = 2 \mu \log \lambda_L + \Lambda \log J.
\end{align}
By imposing that the only not vanishing stress component is the longitudinal one, $ \tau_R = 0 $ implies:
\begin{equation}
\label{legame lambda con ni}
\tau_R = 0  \Longleftrightarrow   \lambda_R = \lambda_L^{-\nu}, \qquad J =  \lambda_R^{2} \lambda_L = \lambda_L^{1-2\nu}\,
\end{equation} 
After simple algebraic manipulations, recalling that the Lamè constants are related to the Young modulus $E$ and the Poisson ratio $\nu$ through the equations $2\mu = \frac{E}{1+\nu}$ and $\Lambda = \frac{\nu E}{(1+\nu)(1-2\nu)}$, the longitudinal Kirchhoff stress takes the form: 
\begin{equation}
\label{tauL ultimo}
\tau_L = E \log \lambda_L.
\end{equation} 
As a consequence, the Cauchy stress \eqref{stresses}$_1$ and the nominal stress \eqref{stresses}$_2$ read as
\begin{equation}
\label{sigmaL}
\sigma_L = \lambda_L^{2\nu - 1} E \log \lambda_L \quad \text{and} \quad P_L = \frac{1}{\lambda_L} E \log \lambda_L.
\end{equation} 
Moreover, a direct integration of \eqref{Hencky SED} over the reference volume $\Omega_0$ of the whole cylindrical element lets to estimate the elastic energy that, by including the constitutive assumption \eqref{Constitutive law}, results 
\begin{equation}
\label{Unuova}
U_H = \int_{\Omega_0} \Psi_H \, dV =\frac{1}{2} \int_{\Omega_0} \tau_L H_L  dV =\frac{1}{2} E A L\left( \log \lambda_L \right) ^{2}.
\end{equation}
where $ A L = Vol(\Omega_0) $.\\
By following the same line of reasoning above, the strain energy function of an element made of a general neo-Hookean material is \cite{holzapfel}:
\begin{equation}\label{neoHSED}
\Psi_{NH}\left( I_1(\textbf{C}), J \right) = \dfrac{c_1}{\beta}\left(  J^{-2 \beta} - 1 \right)  + c_1 \left[ I_1(\textbf{C}) - 3 \right],
\end{equation}
where $ \textbf{C} = \textbf{F}^T\textbf{F} $ is the right Cauchy-Green tensor and the two material constants can be set as $ c_1 = \mu/2 $ and $ \beta = \nu/\left( 1- 2\nu \right) $. By recalling that Kirchhoff stress and Hencky strain are work-conjugate, it results that \cite{Plesek2006}:
\begin{equation}
\tau_i = \dfrac{\partial \Psi_{NH}\left( \textbf{H}\right)}{\partial H_i}, \quad i = \left\lbrace R, L\right\rbrace.
\end{equation}    
Then, by accounting that $\lambda_i = e^{H_i}$, the energy density $\Psi_{NH}$ can be written as an explicit function of the strains $ H_i $ in order to derive the aforementioned stress components. In addition, by imposing $\tau_R = 0$, the same relationship between the transverse and longitudinal stretches \eqref{legame lambda con ni} is obtained, so that the following stress measures can be written in the light of the definitions \eqref{stresses}:
\begin{equation}
\label{tauLNEO}
\tau_L = 2 c_1 \left( \lambda_L^2 -\lambda_L^{-2 \nu} \right), \quad P_L = {\tau_L}{\lambda_L^{-1}} = 2 c_1 \left( \lambda_L -\lambda_L^{-\left( 2 \nu + 1 \right) } \right) ,\quad \sigma_L = J^{-1}\tau_L = 2 c_1 \left( \lambda_L^{ 2 \nu + 1 } - \lambda_L^{-1}\right),
\end{equation} 
while the total energy for the neo-Hookean element reads:
\begin{equation}
\label{energyNEO}
U_{NH} =\int_{\Omega_0} \Psi_{NH}\, dV =  c_1 \left( \lambda_L^2 + \dfrac{\lambda_L^{-2 \nu}}{\nu} - \dfrac{\nu + 1}{\nu} \right) A L.
\end{equation} 
In what follows, we therefore use the equations \eqref{Unuova} and \eqref{energyNEO} (at the end adding the contribution of the bending in compressed bars) as elastic energies from which to derive stress and deformation in cables and struts for both solving the form-finding problem and for obtaining the mechanical response of the \textit{soft} tensegrity under prescribed loading conditions.

\section{Equilibria at symmetry-preserving deformation states}

\subsection{Geometrical relations and equilibria in soft tensegrities}
\begin{figure}[htbp]
\centering
\includegraphics[width=0.5\textwidth]{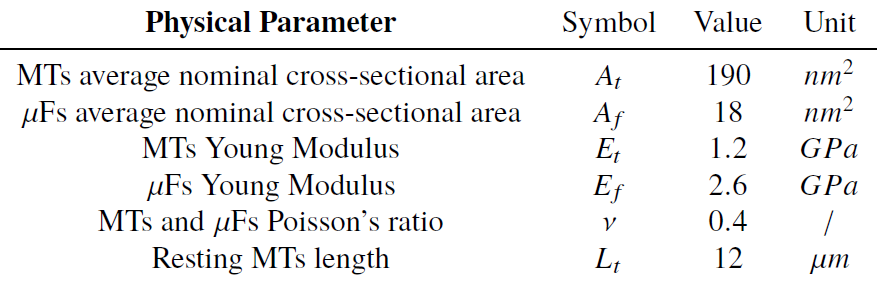}
\caption{Values of geometrical and physical parameters used to simulate the mechanical behavior of tensegrity-based cell cytoskeleton \cite{Gittes_1993}.}
\label{tab.Phisical_parameters}
\end{figure}

\begin{figure} [htbp]
\centering
\includegraphics[width=1\textwidth]{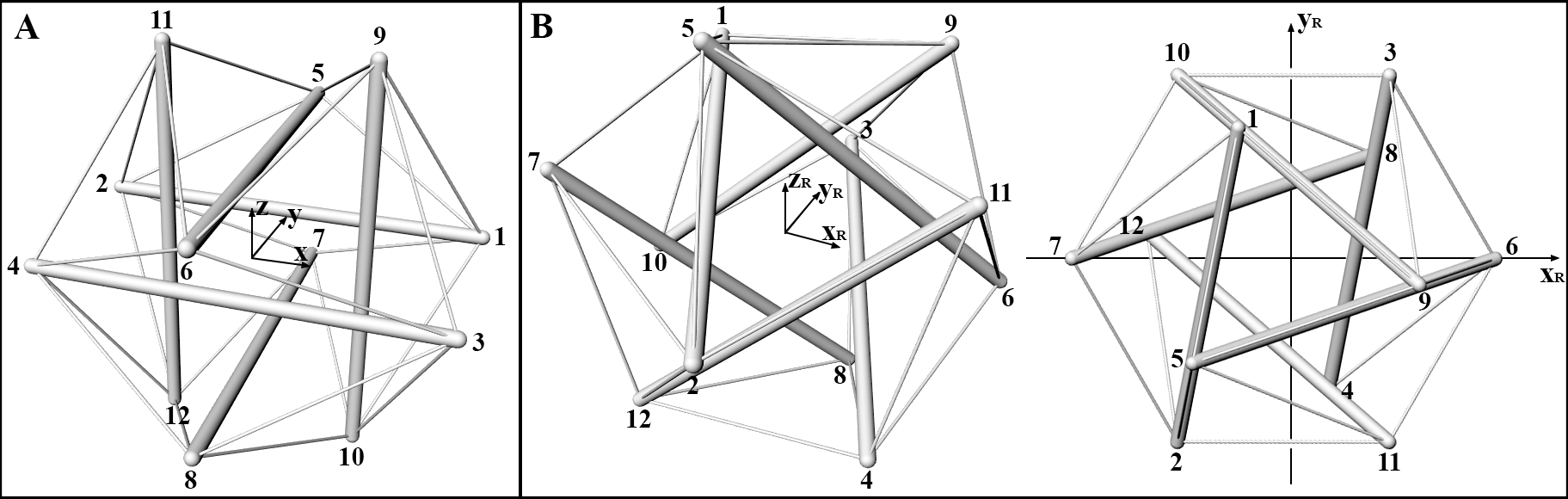}
\caption{\textbf{A)} Perspective view of the $30$-element tensegrity in the Cartesian coordinate system $ \left\lbrace x, y, z \right\rbrace $. \textbf{B)} Three-dimensional (left) and top (right) views of the tensegrity system in the rotated Cartesian reference system $\left\lbrace x_R,y_R,z_R\right\rbrace $ with the latter representation highlighting the geometrical symmetries of the structure.}
\label{fig.30_elem}
\end{figure}

To idealize the cell cytoskeleton, let us consider a $30$-element tensegrity system with a regular icosahedral geometry and let us seek for the pre-stress conditions in cables and struts ensuring self-equilibrium of the whole structure and deformation states compatible with that icosahedral shape (Figure \ref{fig.30_elem}). In such a configuration, the six struts --the cytoskeletal microtubules-- have the same length $ L_t^* $, while the cables --the actin microfilaments-- have length $ L_f^* $. Geometrical arguments and symmetry of the structure require that the actual lengths of struts and cables obey the following equation:
\begin{equation}
\label{relazione_lunghezze}
L_f^* = \dfrac{\sqrt{6}}{4} L_t^*,
\end{equation}
the subscripts $ t $ and $ f $ denoting tubules and filaments, respectively. Also, let us assume that the resting (undeformed) lengths, the initial cross sections and the mechanical properties of all the cables are the same and so happens for the struts, geometrical and constitutive parameters being in the successive calculations referred to those reported in literature and collected in \hyperref[tab.Phisical_parameters]{Table} \ref{tab.Phisical_parameters}. Therein, the elastic moduli and the nominal cross-sectional areas have been chosen according to the experimental data presented by Gittes et al. \cite{Gittes_1993}, while the resting microtubules length has been estimated such that the mean cell diameter (given by $ \sqrt{5} L_t^*/2 $ for the selected geometrical configuration associated to the polyhedral shape of the $30$-element tensegrity) remains always within the range of $ 10-30 \, \mu m $, according to the average sizes observed in many human cells.

From the geometrical point of view, the \textit{topology} of the system is defined by the vertices set, $ V_{TS} $, and the edges set, $ E_{TS} $:
\begin{equation}
V_{TS} = \left\lbrace 1, 2,..., 12\right\rbrace, \qquad E_{TS} = C_{TS} \cup S_{TS}, \,\,\,  C_{TS} = \left\lbrace 1, 2,..., 24\right\rbrace, \,\, S_{TS} = \left\lbrace 25,..., 30\right\rbrace
\end{equation}
where $ C_{TS} $ and $ S_{TS} $ denote the continuous set of cables and the disjoined set of struts, respectively. The \textit{configuration} of the system is instead identified by the vector $ \textbf{\emph{p}} $ containing the $12$ nodal coordinates reported below, written with reference to a Cartesian coordinate system $ \left\lbrace x, y, z \right\rbrace  $,  with the origin placed at the center of the sphere circumscribing all the nodes of the polyhedral tensegrity structure (see Figure \ref{fig.30_elem}\hyperref[fig.30_elem]{A}).
Because of the peculiar polyhedral symmetry exhibited by the $30$-element tensegrity under exam, the coordinates of all nodes can be automatically generated by starting from one of them, by means of rigid transformations; with reference to the Figure \ref{fig.30_elem}, one then has:
\begin{equation}
\label{30_coordinates}
\begin{split}
\textbf{p}_1 = L_t^* \left( \dfrac{1}{2}, \dfrac{1}{4}, 0 \right)^T, \quad \textbf{p}_2 = \textbf{R}_x \textbf{p}_1, \quad \textbf{p}_{5,6} = \textbf{P}_{\pi} \textbf{p}_{1,2}, \quad \textbf{p}_{9,10} = \textbf{P}_{\pi} \textbf{p}_{5,6} = \textbf{P}_{\pi}^2 \textbf{p}_{1,2}, \\
\textbf{p}_{3,4} = \textbf{R}_y \textbf{p}_{1,2}, \quad \textbf{p}_{7,8} = \textbf{P}_{\pi} \textbf{p}_{3,4} = \textbf{P}_{\pi} \textbf{R}_y \textbf{p}_{1,2}, \quad \textbf{p}_{11,12} = \textbf{P}_{\pi} \textbf{p}_{7,8} = \textbf{P}_{\pi}^2 \textbf{R}_y \textbf{p}_{1,2}
\end{split}
\end{equation}
where $ \textbf{P}_{\pi} $ is a permutation matrix, while $ \textbf{R}_x $ and $ \textbf{R}_y $ are reflection matrices with respect to the axes $ x $ and $ y $, respectively given by:

\begin{equation}
\label{permutation_matrix}
\textbf{P}_{\pi}=\left[
\begin{array}{c}
  \textbf{e}_{\pi(1)} \\
  \textbf{e}_{\pi(2)} \\
  \textbf{e}_{\pi(3)} \\
  \end{array}
\right]=\left[
\begin{array}{c}
  \textbf{e}_{3} \\
  \textbf{e}_{1} \\
  \textbf{e}_{2} \\
  \end{array}
\right]=\left[
\begin{array}{ccc}
	0 & 0 & 1 \\
	1 & 0 & 0 \\
	0 & 1 & 0
	\end{array}
	\right],\,\, \textbf{R}_{x}=\textbf{I}-2 \textbf{e}_{1} \otimes \textbf{e}_{1},\,\,
\textbf{R}_{y}=\textbf{I}-2 \textbf{e}_{2} \otimes \textbf{e}_{2}.
\end{equation}
\\

Furthermore, according to the nodal coordinates \eqref{30_coordinates}, it is easy to verify the relationship (\ref{relazione_lunghezze}). As already highlighted, the lengths $ L_t^* $ and $ L_f^* $ refer to the pre-stretched configuration and, therefore, keeping in mind that --at least in self-equilibrated states-- the struts are all compressed and the cables all tensed, they can be related to the respective natural lengths, say $ L_t $ and $ L_f $, through the relationships:
\begin{gather}
\label{L}
L_t^* = \lambda^*_t L_t, \\
\label{l}
L_f^* = \lambda^*_f L_f,
\end{gather}
where $ \lambda^*_t $ and $ \lambda^*_f $ are the homogeneous pre-stretches in struts and cables, respectively, with the inequalities $ 0 < \lambda^*_t \leq 1 $ and $ \lambda^*_f \geq 1 $ which hold true. However, the values of these pre-stretches cannot be independently assigned, since they have to ensure --mediated by the nonlinear elastic laws relating them to the stresses-- equilibrium in the pre-stretched configuration. Hence, in absence of externally applied forces: 
\begin{equation}
\label{equilibrio forze nulle esterne}
\sum_{j} N_{i j}^* \dfrac{\textbf{p}_j-\textbf{p}_i}{\|\textbf{p}_j-\textbf{p}_i\|} = \textbf{0}, \qquad \forall \, i = 1,...,12
\end{equation}
with the summation extended to all nodes $j$ connected to the node $i$ by an element $i$-$j$, $N_{ij}^*$ being the axial force stressing that element. The polyhedral symmetry of the tensegrity module and the hypothesis of equal at rest  lengths of struts and cables, also allow to assume that the pre-stretches and the related pre-stresses have the same values within each group of compressed and tensed elements. By indicating with $ N_t^* $ and $ N_f^* $ the magnitudes of the  axial forces brought by tubules and filaments, the sole equation to be satisfied for preserving symmetry at the equilibrium for each node is: 
\begin{equation}
\label{equil}
N_t^* = - \sqrt{6} N_f^*.
\end{equation}
a result found by imposing the equilibrium along the $z$ direction, the equilibria along the $ x $ and $ y $ axes being automatically ensured by the tensile forces of the four cables converging in any arbitrarily chosen node at the end of a strut. As a consequence, the equation \eqref{equil} alone establishes the relationship that the two pre-stretches defined in \eqref{L} and \eqref{l} must obey, i.e. $N_t^*\left( \lambda_t^* \right) = - \sqrt{6} N_f^*\left( \lambda_f^* \right)$, the forces being obtained by multiplying the resting cross-sectional areas of the element and the nominal stresses $P_L$ coming from one of the two different constitutive laws hypothesized for describing the hyperelastic behavior of cables and struts (see equations \eqref{sigmaL}$_{2}$ and \eqref{tauLNEO}$_{2}$). The symmetry-preserving form-finding problem of the soft-tensegrity is therefore governed by the three compatibility relations given by equations \eqref{relazione_lunghezze}, \eqref{L}, \eqref{l}:
\begin{equation}
\label{eq1}
L_f^* = \dfrac{\sqrt{6}}{4} L_t^*=\dfrac{\sqrt{6}}{4}\lambda^*_t L_t=\lambda^*_f L_f
\end{equation}
to which the equilibrium equation \eqref{equil}, written in terms of pre-stretches accounting for the specific constitutive law, must be added. By following this way, the equations \eqref{eq1} and \eqref{equil} contain six unknowns ($L_f^*, L_t^*, L_f, L_t, \lambda_f^*, \lambda_t^*$) and thus the solution is obtained by treating two of them as parameters. In this case, it seemed appropriate to fix the struts natural length (at the value indicated in Table \ref{tab.Phisical_parameters}) and to parametrically vary the value of the cables pre-stretch in order to evaluate its influence --and consequently the influence of the pre-stress-- on the mechanical response of the structure. As expected, under these conditions, the overall size of the tensegrity module decreases as the pre-stress increases preserving its original shape, the height of the structure being given by $ h = \sqrt{3} L_t \lambda^*_t / 2 $, with $L_t$ fixed and $ \lambda^*_t $ decreasing for increasing $\lambda^*_f $. Obviously, the limit case of inextensible (\textit{rigid}) struts, frequently encountered in the literature \cite{Stamenovic1996, Coughlin1998, Fraternali2015}, is traced back for any possible pre-stressed self-equilibrated state, by increasing the cables pre-stretch and making the elastic modulus of the struts significantly greater than the one of the cables (Table \ref{tab.Phisical_parameters}), say up to the extreme case of rigid struts. In this limit situation, from equation \eqref{eq1}, one in fact has that the relationship $L_f^*/L_t=\sqrt{6}/4$, commonly found in the literature, holds true.\\
With the aim of analyzing the cytoskeleton in self-equilbrium (as in the case of suspended round-shaped cells) and then adherent to the ECM and loaded by external forces, the structure is assumed to stand on a (rigid) substrate and therein anchored through the nodes $ 4, 8 $ and $ 12 $, as shown in Figure \ref{fig.30_elem}\hyperref[fig.30_elem]{B}. From the operational point of view, it is convenient to rotate the reference system in a way that the new $z$-axis intercepts the centers of the equilateral triangles ideally formed by the nodes $ 1-5-9 $ and $ 4-8-12 $. In this new frame of coordinates, referred to as $ \left\lbrace x_R, y_R, z_R \right\rbrace $ system, the nodes $ 1-5-9 $ form the upper triangle, while the lower one is defined by $ 4-8-12 $, whose vertices are thus fully constrained on the rigid substrate (see Figure \ref{fig.30_elem}\hyperref[fig.30_elem]{B}). Also, the new $ \left\lbrace x_R, y_R \right\rbrace $ plane has been oriented in a way that the nodes $ 6 $ and $ 7 $ are identified by null $y_R$ coordinate. As a result, the considered rotation leads to define a new unit vector $\hat{\textbf{z}}_R = \left( \dfrac{1}{\sqrt{3}}, \dfrac{1}{\sqrt{3}}, \dfrac{1}{\sqrt{3}} \right)$, with the other two unit vectors being given by the relations: 
\begin{equation}
\hat{\textbf{x}}_R \cdot \hat{\textbf{z}}_R=0, \,\,\, 
\hat{\textbf{y}}_R \cdot \textbf{p}_6= 0 \, \left( \hat{\textbf{y}}_R \cdot \textbf{p}_7=0\right) ,\,\,\, \hat{\textbf{y}}_R=\hat{\textbf{z}}_R \times \hat{\textbf{x}}_R. 
\end{equation}

This particular choice let to exploit some symmetry properties in studying the mechanical response of the tensegrity experiencing the different deformation regimes (contraction/elongation, torque and shear) as examined below, so minimizing the number of the unknowns and facilitating the seek of the solutions in analytical form.
These intrinsic symmetries can be for example appraised looking at the structure from a top view (Figure \ref{fig.30_elem}\hyperref[fig.30_elem]{B}), and observing a star shaped geometry and a hexagon made up by the nodes of the system, sharing the same center. In what follows, by starting from the analysis of symmetry-preserving deformation modes in response to prescribed boundary conditions, the above mentioned intrinsic symmetries and the peculiar choice of the reference frame will be used.

\subsection{Internal (elastic) energies in symmetry-preserving soft tensegrities}

If we start by excluding that --both in self-equilibrium and under applied loads-- overall tensegrity deformation shapes can deviate from configurations that respect geometrical and load symmetries and also assume that the compressed bars contract without buckling (high bending stiffness of the struts), we can restrict our study to the deformation states here referred as to \textit{symmetry-preserving} ones.\\
Equilibria in pre-stretched configurations and at any stage of deformation induced by external forces can be as usual determined by making the total potential energy stationary, thus minimizing the internal (elastic) energy minus the work done by the applied loads against the corresponding displacements.\\
In order to determine the general form of the internal energy of the polyhedral soft tensegrity, for both the nonlinear (hyperelastic) behaviors to be analyzed, we can start with the case of Hencky's materials \ref{lblb1}: according to (\ref{Unuova}), the energy of each single element (cable or strut) can be written as
\begin{equation}
\label{Usingle}
U_j = \dfrac{1}{2} A_j E_j L_j \left( \log \lambda_j \right)^2, \qquad  j = 1,..., 30,
\end{equation}
where the stretch $ \lambda_j $ is the result of the superposition of two stretches, i.e. the pre-stretch ensuring self-equilibrium of the system and a further elastic stretch due to possible external loads:
\begin{equation}
\label{lambdaj}
\lambda_j = \dfrac{l_j}{L_j} = \dfrac{l_j}{L^*_j} \dfrac{L^*_j}{L_j} = \lambda^*_j \dfrac{l_j}{L^*_j},
\end{equation}
the pre-stretch being indicated with $ \lambda^*_j $, while $ l_j $ is the final length of the $j$-th element, that can be written as a function of the unknown nodal displacements:
\begin{equation}
\label{l_finale}
l_j = \|\textbf{p}'_i(u_i,v_i,w_i)-\textbf{p}'_k(u_k,v_k,w_k)\|.
\end{equation}
In this notation, $i$ and $k$ are the indices of the nodes connected by the $j$-th element, the prime referring to the nodes current coordinates which depend on the nodal displacement vectors $\textbf{u}_i = \left\lbrace u_i,v_i,w_i\right\rbrace, \,\,i=1,...,12$.\\
With reference to the Hencky's model, by taking into account equation \eqref{lambdaj}, \eqref{Usingle} can be readily written as the sum of a term  $U^*_j$ representing the energy of the element due to the sole pre-stretch --which is fixed, once the pre-stretch is provided-- and a term, say $\Delta U_j$, that is the increase of energy due to the applied loads:
\begin{multline}
U_j = \dfrac{1}{2} A_j E_j L_j \left( \log \lambda^*_j \dfrac{l_j}{L^*_j} \right)^2 = \dfrac{1}{2} A_j E_j L_j \left( \log \lambda^*_j + \log\dfrac{l_j}{L^*_j} \right)^2 = \\
= \dfrac{1}{2} A_j E_j L_j \left( \log \lambda^*_j \right)^2 + \dfrac{1}{2} A_j E_j L_j \left[ \left( \log\dfrac{l_j}{L^*_j} \right)^2 + 2 \log \lambda^*_j \log\dfrac{l_j}{L^*_j} \right] = U_j^* + \Delta U_j.
\end{multline}
Finally, the total internal energy of the whole tensegrity can be obtained by summing up the energy aliquots of the single elements:
\begin{equation}\label{HU}
U_H = \sum_j U_j = \sum_j \left[  U_j^* + \Delta U_j \right] = \sum_j  U_j^*  + \sum_j  \Delta U_j = U^* + \Delta U.
\end{equation}
By following the same line of reasoning, the elastic energy for tensegrity systems constituted by neo-Hookean elements can be computed. In this case, however, the total internal energy cannot be additively decomposed and, by accounting \eqref{energyNEO} and \eqref{lambdaj}, one has:    
\begin{multline}\label{NHU}
U_{NH} = \sum_j U_j  = \sum_j  \dfrac{\mu_j}{2} \left( \lambda_j^2 + \dfrac{\lambda_j^{-2 \nu_j}}{\nu_j} - \dfrac{\nu_j + 1}{\nu_j} \right) A_j L_j =\\
= \sum_j  \dfrac{\mu_j}{2} \left( \left(\lambda^*_j \dfrac{l_j}{L^*_j}\right)^2 + \left({\left(\lambda_j^* \dfrac{l_j}{L^*_j}\right)^{-2 \nu_j}} - \nu_j - 1\right){\nu_j}^{-1} \right) A_j L_j.
\end{multline}

\subsubsection{Form-finding and energy storing in cell cytoskeleton}
\begin{figure} [htbp]
\centering
\includegraphics[width=1\textwidth]{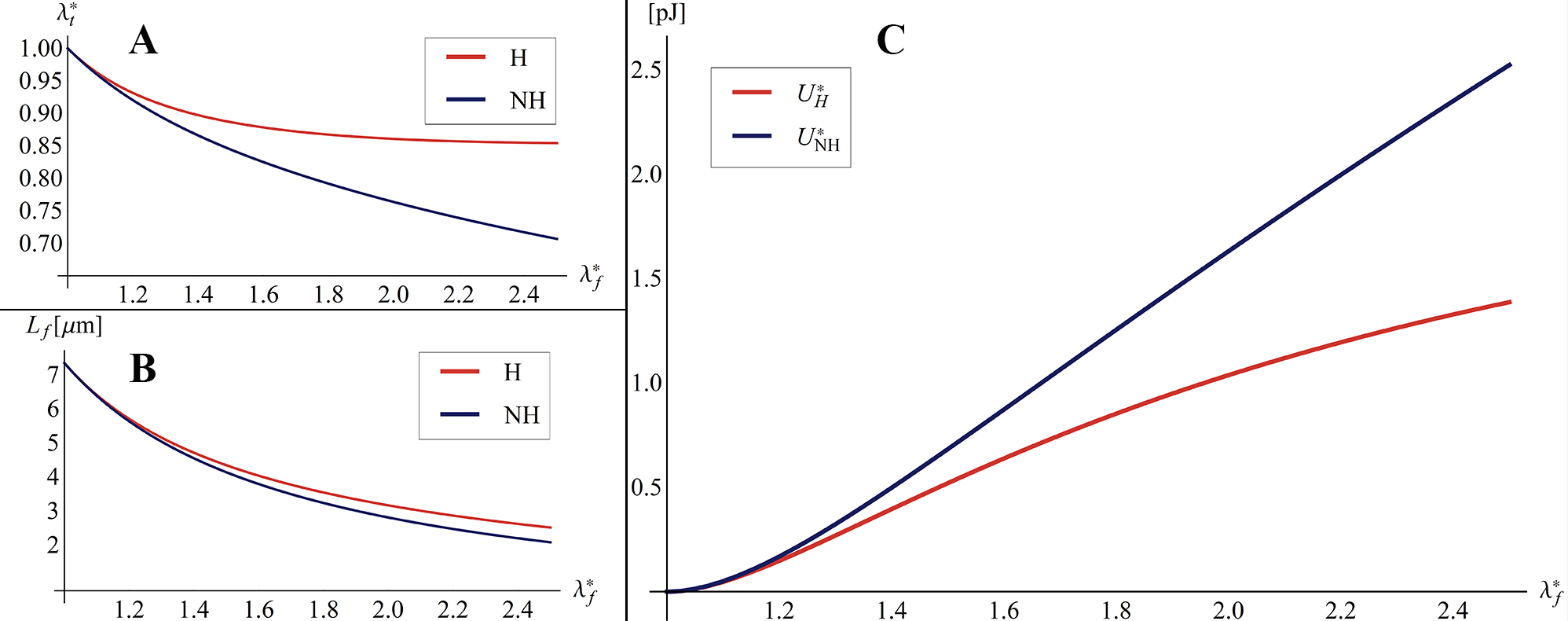}
\caption{Change of \textbf{A}) microtubules's prestretch, \textbf{B}) microfilaments' rest length and \textbf{C}) energy stored by the cell-tensegrity at the self-equilibrated pre-stretched state, while increasing the filaments' pre-stretch values from $1$ to $2.5$, both for Hencky's (H) and neo-Hookean (NH) models.}
\label{fig.formfind}
\end{figure}

As first, we can analyze the case of absence of external loads, say the form-finding problem of the idealized cytoskeleton. Equilibrium equation \eqref{equil} can be particularized for the cases of elements obeying the Hencky and the neo-Hookean laws, by also employing the compatibility relationships \eqref{eq1}. Then, by considering the nominal stresses \eqref{sigmaL}$_2$ and \eqref{tauLNEO}$_2$, one finds that the prestretch in microtubules is driven by that in microfilaments according to the following balance relations:
\begin{equation}
\begin{split}
&E_tA_t\dfrac{\log\lambda_t^*}{\lambda_t^*}=-\sqrt{6}E_fA_f\dfrac{\log\lambda_f^*}{\lambda_f^*}\quad \text{for Hencky-type elements, while}\\
&\mu_tA_t\left[ \lambda_t^*-\left( \lambda_t^*\right) ^{-\left(2\nu_t+1\right)}\right] =-\sqrt{6}\mu_fA_f\left[ \lambda_f^*-\left( \lambda_f^*\right) ^{-\left(2\nu_f+1\right)}\right]\quad \text{for neo-Hookean ones,}
\end{split}
\end{equation} 
that, by introducing the values of the parameters given in Table \ref{tab.Phisical_parameters}, provide the results shown in Fig.\ref{fig.formfind}\hyperref[fig.formfind]{A} for $\lambda_t^*$ and in Fig.\ref{fig.formfind}\hyperref[fig.formfind]{B} for compatible rest length $L_f^*$ of microfilaments, obtained by virtue of eq.\eqref{eq1}. In this purely pre-stretched/pre-stressed self equilibrated state --which preserves the polyhedral shape of the tensegrity and thus can be seen as the configuration assumed by a suspended round cell-- the  expressions of the energy stored by the tensegrity structure take the forms respectively given by eq.\eqref{HU}, in case of Hencky's model, and by eq.\eqref{NHU}, in case of neo-Hookean elements:
\begin{equation}
\begin{split}
&U_H^*=3A_tE_tL_t\left(\log\lambda_t^*\right)^2+12A_fE_fL_f\left(\log\lambda_f^*\right)^2\quad\text{and}\\
&U_{NH}^*=3 \mu_tA_tL_t\left[\left(\lambda_t^*\right)^2+\dfrac{\left(\lambda_t^*\right)^{-2\nu_t}}{\nu_t}-\dfrac{\nu_t+1}{\nu_t}\right]+12 \mu_fA_fL_f\left[\left(\lambda_f^*\right)^2+\dfrac{\left(\lambda_f^*\right)^{-2\nu_f}}{\nu_f}-\dfrac{\nu_f+1}{\nu_f}\right]
\end{split},
\end{equation}
the amount of stored energy increasing with $\lambda_f^*$ as shown in Fig.\ref{fig.formfind}\hyperref[fig.formfind]{C} and --in the case of soft-bar tensegrity-- resulting also accompanied by an overall cell shrinking (see Fig.\ref{fig.formfind}\hyperref[fig.formfind]{B}). 
The results show that the introduction of hyperelasticity could more faithfully reflect the nonlinear way of a cell to accumulate elastic energy through the pre-stress of its protein filaments, here predicted to be of the order of a few units of $pJ$. Confirming this quantitative result is however not so easy. We know that the main source of energy is allocated in cell proteins and other organic complexes and most part of it is spent to form molecules. As a function of the releasing times and of the provision needs the cell energy is stored at long term in lipids (e.g. triglycerides and adipocytes), at short term --say about 24 hour supply-- in the liver (glycocen) and for immediate use as Adenosine TriPhosphate (ATP), the (chemical) energy currency of all living cells, generated by cellular respiration, stored in the bonds that held the atoms of molecule together and released by breaking into ADP (adenosine diphosphate) and inorganic phosphate, with the reaction catalyzed by ATPase enzymes. Despite all these mechanisms are known, obtaining a reliable estimate of the energy storage and of the energy rate production in human cells still remains a tricky task, these values strongly depending on the very different compositions, sizes, growth conditions and functions characterizing each cell line. It would be in fact sufficient thinking that, for instance, fibroblasts are significantly more active than the average human cell, thus requiring higher energy reserves to be used. Moreover, the major oxygen-consuming processes --e.g. protein synthesis, $Na^+/K^+$ ATPase (responsible of maintaining the resting electric potential in cells) $Ca^{2+}$ and actinomyosin ATPase (that drives muscle cells)-- are found with extremely variable percentages in liver, heart, brain, skeletal muscle cells and other human tissues \cite{Rolfe_1997,Bernstein_2003}. These differences might therefore call into question the accuracy of any estimate of stored energy per cell if one does not admit possible discrepancies of two (or more) order of magnitude when the average values are compared with experimental data related to a specific cell line.\\
However, by using the rule of thumb and starting from a caloric intake of about $2000$ kcal per day in an adult human of medium build, rough calculations lead to estimate an overall heat production at a rate of about one hundred watts (100 joules per second), corresponding to a few units of pico-watts per cell, if we consider about $30$ billion ($3\times10^{13}$) of cells which populate the human body \cite{natgeo}. Nevertheless, as already pointed out above, bottom-up analyses may conduce up to two order of magnitude greater values if selected cells are taken into account.\\
Aware of this variability of data, the order of magnitude of the elastic energy storage predicted by our cellular soft-strut tensegrity unit seems to be however consistent with some estimates supported by experimental findings.
By referring to \cite{energy,Hooper_2008} for a more detailed discussion on the molecular basis of contraction and regulation in vertebrate and invertebrate muscles, it can be for example shown that the elastic energy storage in myofilament lattice depends on sarcomere length and, by comparing the energy input due to the consumption of ATP to the energy stored across all filaments and cross-bridges, values of energy stored by a single sarcomere were estimated not to exceed $1000$ $pN \times nm$. By considering that a muscle fiber may contain about $10^5$ sarcomeres, we can therefore calculate a stored elastic energy of about $10^{-1}$ $pJ$, that is in full agreement with the elastic energy accumulated by our tensegrity model when the pre-stretches in the filaments fall between $1.1$ and $1.2$, these values being consistent with the actual average strain ranging from $10\%$ to $20\%$ in a muscle fiber (see Fig. \ref{fig.formfind}\hyperref[fig.formfind]{C}).\\
A further confirmation of the capability of the proposed soft-strut tensegrity model to predict the order of magnitude of the energy storage in a cell can be also found by directly making reference to ATP. In fact, it can be demonstrated that in many eukaryotic cells, motility is driven by dynamic actin polymerization at a steady state cost of about $1$ ATP hydrolysis per polymerizing actin monomer \cite{Pollard_2003,Abraham_1999}. Comparative studies show that an energy rate of $4 \times 10^5$ ATP/s, associated to about $4000$ filaments, is required to power cell movement \cite{Svitkina_1997}. On the other hand, the rule of thumb involving Gibbs free energy change due to ATP hydrolysis \cite{ROSING1972275} and calculations of forces due to a molecular motor allow to predict that it would exert a force of roughly $5$ $pN$ \cite{PhyBioCell2012} over a $10$ $nm$ \cite{cellbiobook}, then doing a work of order $50$ $pN \times nm$ which requires slightly more than $10$ $k_B \times T$ of energy ($k_B$ being the Boltzmann constant and $T$ the absolute temperature), well within the range of what a single ATP can deliver \cite{dill2012molecular}. Therefore, by converting the energy rate of $4 \times 10^5$ ATP/s in pico-joules per second, then multiplying this result for $8.64 \times 10^4$ seconds a day and dividing it by $36 / 4000$ (the ratio between the tensegrity elements and the total number of filaments on which the above energy amount has been estimated), one finally obtains about $15$ $pJ$, consistent with the amount of elastic energy stored by the cellular tensegrity model, that hence would transform about $10\%$ of the total chemical ATP in elastic energy.\\
Finally, beyond any quantitative confirmation, it is extremely worth noticing that the \textit{soft}-strut tensegrity model confers the cytoskeleton the capability to combine energy storing with cell size modulation, by so adapting the tensegrity paradigm to the actual peculiar behavior of living cells. In fact, while pre-stress is used by cells for regulating many biochemical signals and \textit{ad hoc} releasing energy for adhering to ECM, migrating and reorienting over substrates and governing some cell-cell interactions, size tuning and shrinking --absent in classical \textit{rigid}-strut tensegrity models-- are instead crucial features exploited by round as well as deforming cells for spreading, to sneak into blood vessels and to overcome micro-channel obstructions \cite{Shelby_2003}, and by cancer cells to gain capabilities to gatecrash in remote districts so promoting metastasis \cite{Plodinec_2012,Pachenari_2014,Ketene_2012}.

\subsection{Symmetric responses of cellular soft tensegrity under applied loads}
Here, we consider that cells, by starting from their self-equilibrated pre-stressed configurations, deform under the action of external loads applied in terms of prescribed displacements at the three nodes placed at the top of the tensegrity structure, then impeding any degree of freedom at the corresponding three nodes at the basis of the system. In this way, the total potential energy coincides with the internal energy and the equilibrium is found by making stationary $U$ with respect to the vector collecting all the unknown nodal displacements components, say $\tilde{\textbf{u}}_i$. The problem to be solved so reduces to the following minimization:
\begin{equation}
\label{minimprob}
\tilde{\textbf{u}}_i:\,\,\,\min_{\tilde{\textbf{u}}_i}U\,\, \Leftrightarrow \,\, \partial_{\tilde{\textbf{u}}_i}U=\textbf{0},\,\,\text{with}\,\,\textit{\textbf{H}}_{\tilde{\textbf{u}}_i}\left(U\right) \, \text{positive definite}\,\forall\,i\,\in \mathbb{I}\subset\mathbb{N}
\end{equation}
where $\textit{\textbf{H}}_{\tilde{\textbf{u}}_i}\left(U\right)$ is the Hessian of $U$ whose derivatives are calculated with respect to $\tilde{\textbf{u}}_i$ and $\mathbb{I}$ denotes the subset of the natural numbers collecting the indexes $i$ such that the related nodes have at least one degree of freedom. Therefore, by recalling the expressions of $U$ given in \eqref{HU} and \eqref{NHU}, the following systems of nonlinear equations have to be solved to have equilibrium, in the cases of Hencky and neo-Hookean elements, respectively:
\begin{equation}
\begin{split}
&\partial_{\tilde{\textbf{u}}_i}U_H=\sum_j E_j A_j \dfrac{L_j}{l_j^2} \log \dfrac{l_j}{L_j} \left(\textbf{p}'_i-\textbf{p}'_k\right)=\textbf{0},\,\forall\,i\,\in \mathbb{I} \quad \text{and}\\
&\partial_{\tilde{\textbf{u}}_i}U_{NH}=\sum_j\dfrac{\mu_j A_j}{l_j}\left[\dfrac{l_j}{L_j}-\left( \dfrac{L_j}{l_j}\right)^{2\nu_j+1} \right]\left(\textbf{p}'_i-\textbf{p}'_k\right)=\textbf{0},\,\forall\,i\,\in \mathbb{I},
\end{split}
\end{equation}
with the summation extended to all the elements $j$ having one endpoint in the $i$-th node.\\
The non-algebraic and nonlinear structure of both the systems did not allow to solve them in closed-form. We thus solved the minimization problems numerically, by exploiting the Newton's method implemented by the function \textit{FindMinimum} provided by the commercial code \emph{Mathematica}\textsuperscript\textregistered \cite{MathematicaProgram} and double checking the results through an \textit{ad hoc} algorithm based on a random procedure.
It involved the definition of a starting Gaussian-type distribution $\mathfrak{N}$ with mean $\upsilon=0$ and standard deviation $\varpi$ proportional to the value of the prescribed displacement ($p.d.$) according to
\begin{equation}
\label{distribution}
\mathfrak{N}(\upsilon, \varpi) = \mathfrak{N}\left( 0\,, \dfrac{4}{10}(\vert p.d. \vert + 0.0001)\right) 
\end{equation}
from which the values to be initially assigned to the unknown displacements can be extracted from the values around zero, adding up the constant $ 0.0001$ to ensure that at $p.d. = 0$, $\varpi>0$. Successively, random values were extracted from the distribution \eqref{distribution} and assigned to the unknown displacements, by calculating the corresponding energy. This procedure was thus repeated a number of times much greater (at least three order of magnitude) than the number of displacements to be determined (depending on the type of test to simulate) and then, among all the energy values obtained, the minimum was extracted, together with the values of the unknown displacements in correspondence of which the minimum occurred. These values were then used as means of new Gaussian distributions --one for each displacement-- whose standard deviation was halved than before. The random minimization step was so repeated and the values extracted from the distributions hence found to be closer to the minimum point. The operation was therefore iterated with, in particular, five repetitions. The double check was then made by comparing the outcomes of the random procedure with those obtained by applying the function \textit{FindMinimum} provided by the software \emph{Mathematica}\textsuperscript\textregistered and the very good agreement between the two outputs was finally used as a measure of the reliability of the obtained results.

\subsubsection{Crushing and stretching of cells: contraction and elongation}
Let us start by analyzing the case of a cellular (soft-strut) tensegrity which simply contracts or elongates as downward or upward uniform vertical displacements are prescribed on the upper nodes $1, 5, 9$ of the structure (see Figure \ref{fig.30_elem}\hyperref[fig.30_elem]{B}), while the nodes $4, 8, 12$ at the basis are constrained, say anchored to the substrate. The unknowns of this problem are the Cartesian components of the displacements of the nodes belonging to the middle hexagon and the sole in-plane components of the upper nodes. To further reduce the number of unknowns, the symmetry of the structure and of the expected deformation can be both exploited to impose that the nodes forming the above mentioned middle hexagon (as seen from a top view), placed at the same height, share the same vertical displacement to preserve the aforementioned symmetry. Therefore, the $z_R$-components of the displacement of the nodes $3, 7, 11$ and of the nodes $2, 6, 10$ have to satisfy the following equations and can be conveniently re-baptized as:
\begin{align}
\label{vertical_disp}
\notag &w_{HT} := w_3 = w_7 = w_{11}, \\
&w_{HB} := w_2 = w_6 = w_{10}, 
\end{align}
where the subscript $HT$ refers to the nodes belonging to the middle hexagon at higher height, while the subscript $HB$ is used to indicate the lower nodes. 
Additionally, symmetry implies that the radial and tangential displacements take the same values separately for the sets of nodes $3, 7, 11$ and $2, 6, 10$ of the hexagon, respectively, this holding true for nodes $1, 5, 9$, too. This means that a local two-dimensional reference system lying in the $\left\lbrace x_R, y_R \right\rbrace $ plane can be introduced for each of these nodes, rotated in such a way that the new ordinate axis lies in the radial direction. Then, by indicating in the local frames the common radial and tangential displacements of the $i^{th}$ node with $d_{ri}$ and $d_{ti}$, the displacements $u_i$ and $v_i$ of such nodes along the axes $x_R$ and $y_R$ can be determined as follows:
\begin{equation}
\label{u_i v_i matriciale}
\left[
\begin{array}{c}
	u_i \\
	v_i
	\end{array}
	\right] = \textbf{R}(\alpha_i)\left[
\begin{array}{c}
	d_{ti} \\
	d_{ri}
	\end{array}
	\right], 
	\quad \text{with}\quad
	\textbf{R}(\alpha_i) := \left[
\begin{array}{cc}
	\cos\alpha_i & \sin\alpha_i \\
	-\sin\alpha_i & \cos\alpha_i
	\end{array} 
	\right],
\end{equation}
where $d_{ri}:=d_{rUT}$ and $d_{ti}:=d_{tUT}$ for $i=\{1,5,9\}$, $d_{ri}:=d_{rHT}$ and $d_{ti}:=d_{tHT}$ for $i=\{3, 7, 11\}$, $d_{ri}:=d_{rHB}$ and $d_{ti}:=d_{tHB}$ for $i=\{2, 6, 10\}$, while $ \textbf{R}(\alpha_i) $ is the clockwise rotation matrix defined in \eqref{u_i v_i matriciale}$_{2}$ as a function of the angle $ \alpha_i $, defined with respect to the axis $y_R$ and depending on the position of the specific node in the Cartesian frame:
\begin{align}
\label{angles}
\notag &\alpha_1 = 2\pi - \arccos \left( \dfrac{5}{2 \sqrt{7}} \right) , & &\alpha_5 = 2 \pi - \arccos \left( - \dfrac{2}{\sqrt{7}} \right) , & &\alpha_9 = \arccos  \left(-\dfrac{1}{2 \sqrt{7}} \right), \\
\notag &\alpha_2 = \dfrac{7}{6}\pi, & &\alpha_3 = \dfrac{\pi}{6}, & &\alpha_6 = \dfrac{\pi}{2}, \\ 
&\alpha_7 = \dfrac{3}{2}\pi, & &\alpha_{10} = \dfrac{11}{6}\pi, & &\alpha_{11} = \dfrac{5}{6}\pi.
\end{align}

\begin{figure} [htbp]
\centering
\includegraphics[width=1\textwidth]{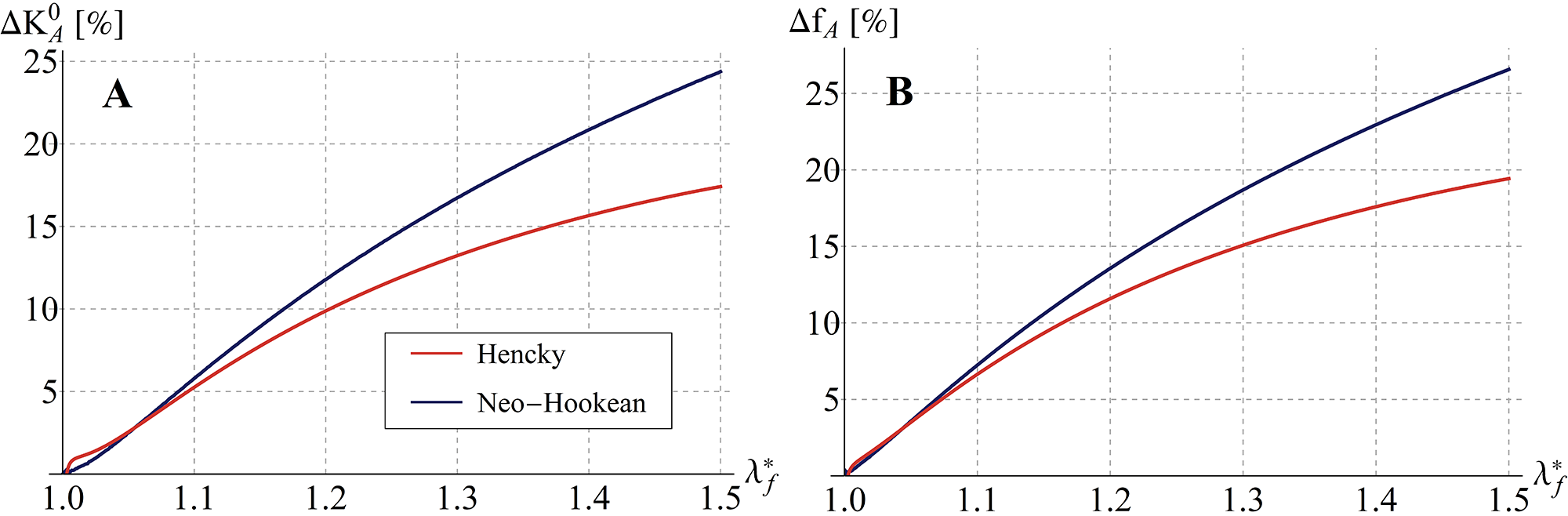}
\caption{Percentage difference in terms of \textbf{A}) initial (tangent) stiffness and \textbf{B}) proper frequency obtained by comparing the presented soft tensegrity with a standard rigid-strut model, while prescribing growing filaments' pre-stretch, for a contraction/elongation loading type, under both the assumptions of Hencky's and neo-Hookean constitutive law.}
\label{fig.delta}
\end{figure}

Under these considerations, the number of unknowns reduces to eight, namely $d_{rUT}$, $d_{tUT}$, $d_{rHT}$, $d_{tHT}$, $d_{rHB}$, $d_{tHB}$, $w_{HT}$, $w_{HB}$, while the vertical displacement $W$ of the upper equilateral triangle is prescribed and the displacement components of the lower nodes set to be zero.\\
A view of the tensegrity deformation process is shown in Figure \ref{fig.DefPlot}, for cables' pre-stretch equal to $1.1$ and a prescribed displacement up to $\pm h/3$ for both elongation and contraction. The results, obtained by means of Hencky and neo-Hookean models, did not exhibit significant differences in terms of deformed configurations and therefore an unique plot was reported. During contraction, the tensegrity rotates counterclockwise and expands laterally, while clockwise rotation and lateral contraction occur in elongation. Noteworthy, this peculiar coupling of torsional rotation with axial and lateral deformations shown by the tensegrity undergoing contraction/elongation may have interesting implications in the analysis of some collective behaviors of cells. In fact, gastrulation during wound healing \cite{Wolgemuth2011}, as well as the experimentally observed geometrical confinement of cells into well-defined circles, that induces a persistent, coordinated and synchronized rotation of cells \cite{Doxzen2013} during their collective migration, are nowadays modeled through \textit{top-down} macroscopic continuum descriptions based on the nematic liquid crystals theory by thus \textit{a priori} imposing the peculiar kinematics. As a consequence, tensegrity models, that intrinsically relate torsion to lateral deformation, could helpfully contribute to construct, for example via homogenization, a rationale \textit{bottom-up} way for deriving enriched continua for interpreting the above mentioned phenomena.\\
Other relevant results are illustrated in the Figures \ref{fig.K}\hyperref[fig.K]{A-B}, that show the overall cell stiffness $K_{A}$ of the structure as a function of the equivalent strain $\varepsilon_{eq}$ (also reported in terms of nominal stress $P_{A}$ versus strain in the insets), when different values of the cables' pre-stretch are considered, for both the cases of cytoskeletal elements obeying Hencky and neo-Hookean laws. More in detail, the nominal stress $P_{A}$ is here defined as the ratio between the equivalent reaction force $F_{A}$ --obtained as derivative of the internal energy with respect to the applied displacement $W$-- and the area of the pre-stretched upper equilateral triangle 
$ A_{tr} = 3 \sqrt{3} L_t^2 \left( \lambda^*_t\right) ^2 / 32 $, that is $ P_{A}:=F_{A}/A_{tr}$, with $F_{A}={\partial U }/{\partial W}$. Moreover, the equivalent stiffness $ K_{A}$ is defined as $ K_{A} := {\partial P_{A} }/{\partial \varepsilon_{eq}} $, where $ \varepsilon_{eq} $ is the ratio between the prescribed displacement and the height of the pre-stretched tensegrity, i.e. $ \varepsilon_{eq} := W/h $. Figure \ref{fig.K}\hyperref[fig.K]{A} shows that the Hencky tensegrity exhibits a hardening, both in contraction and elongation, as the deformation level is increased at low values of $\lambda_f^*$, while a stiffening in elongation and a softening in contraction are registered for higer values of $\lambda_f^*$, with a trend that inverts this behavior as $\lambda_f^*$ grows. The case of neo-Hookean tensegrity (figure \ref{fig.K}\hyperref[fig.K]{B} also provides a hardening by increasing the deformation level, both in contraction and elongation and for low values of $\lambda_f^*$, exhibiting instead always a stiffness increase in elongation and a stiffness decrease in contraction, for higer values of $\lambda_f^*$, somehow qualitatively resembling the results very recently obtained by Fraternali et al. \cite{Fraternali2015} for a simpler three-(rigid)strut tensegrity with cables obeying the de Saint Venant-Kirchhoff law. Note that, in Figures \ref{fig.K}, we report tracts of the curves in grey to provide theoretical extrapolations corresponding to branches which \textit{de facto} cannot be followed, since they would refer to cables bearing compression \footnote{Rigorously speaking, the tracts of the curves in grey indicate that at least one cable --or more likely a set of them-- would undergo compression, this implying, in most of the cases examined, that the whole equilibrium is compromised or simply that the tensegrity should switch on other possible configurations no longer preserving the symmetry, in order to explore eventual different equilibria states. These possible alternative states, that could involve contraction and buckling of struts and/or global deviation of the deformed system from regular shapes, are investigated in the next section of the present work, just to analyze what happens in cases of symmetry losing. However, it should be emphasized that asymmetrical configurations are not a "safe harbor" where to find equilbria otherwise impossible. Also, they could compete with symmetry-preserving configurations in minimizing the tensegrity energy --in pre-stressed or under external loads-- also if symmetry-preserving equilibrium states were possible.}. As a matter of fact, such a case is also incompatible in cells where cable-like cytoskeleton contractile actin microfilaments absorb only tensile forces and the compressive stresses are supported by microtubules \cite{Stamenovic_2001,Wang2001MT,Pathak2012}.\\
It is worth to highlight that the two (neo-Hookean and Hencky) tensegrities exhibit different behaviors in elongation ($ \varepsilon_{eq} > 0 $) and contraction ($ \varepsilon_{eq} < 0 $), as well as very different trends for varying values of the pre-stretch. The response of the structure in fact depends on the harboring pre-stress level which, in turn, governs the initial (tangent) stiffness of the tensegrity system, as shown in Figure \ref{fig.K0}\hyperref[fig.K0]{A}. In particular, both Hencky and neo-Hookean models exhibit a non-zero tangent stiffness at early stage of contraction/elongation if a not vanishing pre-stress is present, the magnitude of this initial stiffness being closely related to the pre-stress value determined by the hyperelastic law chosen for the elements. However, the cell initial (tangent) stiffness is significantly different in the two cases considered  (Figure \ref{fig.K0}\hyperref[fig.K0]): for the neo-Hookean case, it monotonically increases as the pre-stretch in the cables increases, as actually found in some theoretical predictions \cite{Wang1994} and experimental results \cite{Wang_2001}, while --for the Hencky model-- the initial stiffness shows a counterintuitive decreasing path from a selected threshold similar to that found by Coughlin and Stamenovic in their "round" tensegrity model comprising rigid struts \cite{Stamenovic1996,Coughlin1998}, that however seems to have not been experimentally observed so far.\\
Finally, from the quantitative point of view, we noticed that the values of the overall cell stiffness obtained by modeling the cytoskeleton as a soft-strut tensegrity, gave values of the order of magnitude of about $10^2-10^3 \, Pa$, spanning over a reasonable wide range of prestress, in line with the most commonly ascertained values of stiffness measured in the literature through several experimental techniques, for different healthy and cancer cell lines \cite{Stamenovic_2002,Fraldi_2015}.\\
By way of example, it can be useful to compare the initial (tangent) stiffness evaluated for the proposed soft tensegrity model with that provided by a classical rigid-strut one. To this end, Figure \ref{fig.delta}\hyperref[fig.delta]{A} shows that, as the system pre-stress grows, differences in stiffness increase, by reaching percentages up to about $25\%$ and $17\%$ --respectively for neo-Hookean and Hencky's constitutive laws-- as the filaments' prestretch tends to $1.5$. Coherently, a similar result in terms of proper frequencies is found by comparing standard and soft-strut models when oscillating by contracting/elongating around the tensegrity pre-stressed equilibrium position. By solving the \textit{small-on-large} problem, we in fact determined the proper frequencies as $f_A=\sqrt{\left({\partial F_{A} }/{\partial W}\right) M_{cell}^{-1}}$, $M_{cell}$ representing a rough estimate of the cell mass obtained by multiplying the volume of an equivalent sphere circumscribing the structure for the cytosol density, which is about the one of the water \cite{Fraldi_2015} (see Figure \ref{fig.delta}\hyperref[fig.delta]{B}).

\subsubsection{Shearing of the cell cytoskeleton}

Cells experience shear stresses in many \textit{in vivo} situations. Osteocytes inhabiting the lacunae across osteon lamellae regulate the bone mineral unit (BMU) activity by sensing solid and fluid-induced shear stresses, so mediating the mechanical signaling to orchestrate the cell mechanobiology and the turnover of osteoblasts and osteoclasts \cite{Han16689}. Shear stresses are also sensed by endothelial cells forming the monolayer of the intima, the innermost tunica of an artery or a vein, the blood flow continuously stimulating them through tangential forces that are at the basis of important biomechanical processes \cite{Fungbook}, including vessel growth and remodeling \cite{Holzapfel2000art,Nappi_2015}.\\
To simulate shear loading on a cell the tensegrity model is then constrained at its basis and subjected to an uniform displacement in the $ \left\lbrace x_R, y_R \right\rbrace $ plane, prescribed to the upper three nodes of the system so that:
\begin{equation}
\begin{split}
U &:= u_1 = u_5 = u_9, \\
V &:= v_1 = v_5 = v_9, \\
w_{UT} &:= w_1 = w_5 = w_9 ,
\end{split}
\label{vert_disp2}
\end{equation} 
where $ w_{UT} $ is unknown, while $ U $ and $ V $ are respectively assigned along the axes $ x_R $ and $ y_R $ and set equal to:
\begin{equation}
\begin{split}
U &= D_S \cos \beta,\\
V &= D_S \sin \beta, 
\end{split}
\end{equation} 
where the displacement magnitude $ D_S $ in the $ \left\lbrace x_R, y_R \right\rbrace $ plane and its direction with respect to the $x_R$-axis, said $ \beta $, are data. The number of unknowns for the case at hand is then $19$. Differently from the previous contraction-elongation test, in case of shear this number cannot be further reduced, since we have no longer axial symmetry. The unknowns of the problem are thus $ w_{UT}$, $ w_2$, $w_3$, $ w_6$, $w_7$, $ w_{10}$, $w_{11}$, $u_2$, $v_2$, $u_3$, $v_3$, $u_6$, $v_6$, $u_7$, $v_7$, $u_{10}$, $v_{10}$, $u_{11}$ and $v_{11} $, as usual the subscript referring to the node number and $u$, $v$ and $w$ denoting the corresponding displacement components parallel to the axes of the Cartesian reference frame.\\
The results, in terms of overall cell deformation, are shown in Figure \ref{fig.DefPlot}, for $\lambda_f^* = 1.1$, $\beta = \pi/2-\alpha_8 = \arccos \sqrt{{3}/{7}} $ and for a prescribed shear displacement up to $L^*_f$. We did not report sensitivity analyses by varying the value of the angle $\beta$, this being pointless since the geometrical symmetry of the structure would imply a periodicity of the shear response with period $2\pi/3$.
The cell equivalent shear modulus $K_S:={\partial P_{S}}/{\partial \gamma_{eq}}$ and the nominal stress $ P_{S} $ are both represented as functions of the equivalent shear strain --here defined as $ \gamma_{eq} := D_S/h $-- for different values of the pre-stretch in the cables and for both Hencky and neo-Hookean laws, as illustrated in Figures \ref{fig.K}\hyperref[fig.K]{C-D}, respectively. The plots show a decrease of the shear stiffness as the strain level increases. As expected, also in this case the value of the pre-stretch $\lambda^*_f $ strongly affects the initial shear modulus $K_{S}^0$,  that behaves very differently for the two hyperelastic models analyzed, exhibiting a stiffness decrease when the Hencky's model is adopted --in analogy to the case of contraction/elongation-- and an almost linear hardening in the neo-Hookean case, that is still in agreement with experimental and previous theoretical results \cite{Stamenovic2000, Wang_2001, Stamenovic} (see Figure \ref{fig.K0}\hyperref[fig.K0]{B}).\\
It is worth noticing that, in the present case, the results demonstrate that the curve $K_{S}$ plotted against the equivalent shear strain cannot exhibit even a valid (reliable) tract if $\lambda_f^*=1$ (it is entirely grey in Figure \ref{fig.K}\hyperref[fig.K]{C-D}), not even for $K_{S}^0$ (that is zero at $\lambda_f^*=1$). This is since, without an initial pre-stretch, some cables immediately would experience a not admissible compressive stress state, also leading to loss of equilibrium for the entire system at the early stage of shear.

\subsubsection{Overall torque of cells}

\begin{figure} [htbp]
\centering
\includegraphics[width=1\textwidth]{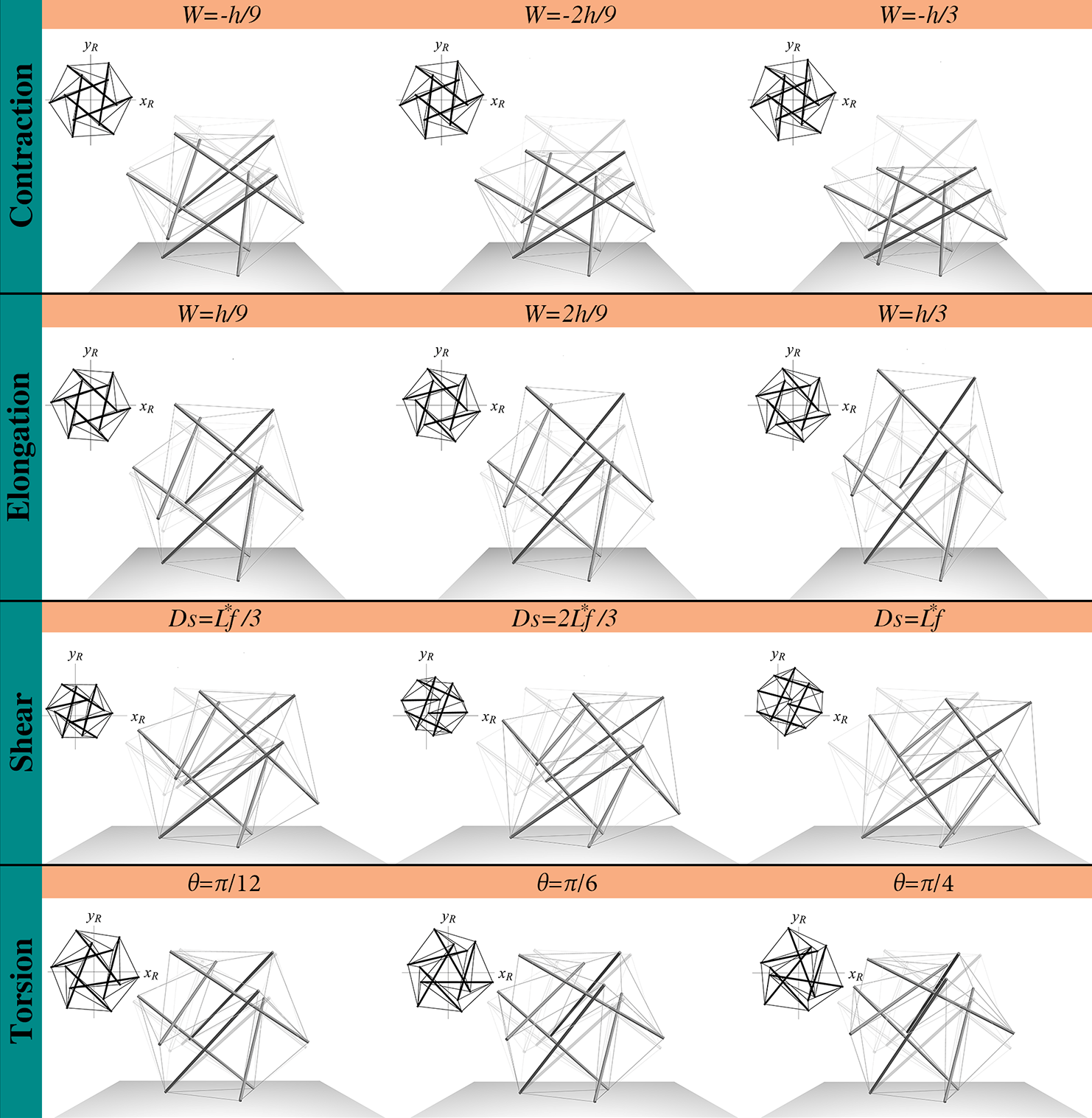}
\caption{3D front view --and top view (at the top left of each image)-- of the deformation sequences of the cellular tensegrity model under the action of the prescribed mechanical conditions, for different values of the assigned displacements (contraction/elongation and shear) and the rotation angle (torsion). Here, $h$ and $L_f^*$ are the tensegrity height and the cables length in the pre-sretched configuration, respectively, and in all the cases the value of cables pre-stretch $\lambda_f^*=1.1$ is set. Light-colored on the background of each image shows the pre-stretched configuration. The values of the parameters used for the analysis are reported in Table \ref{tab.Phisical_parameters}.}
\label{fig.DefPlot}
\end{figure}

\begin{figure} [htbp]
\centering
{\includegraphics[width=1\textwidth]{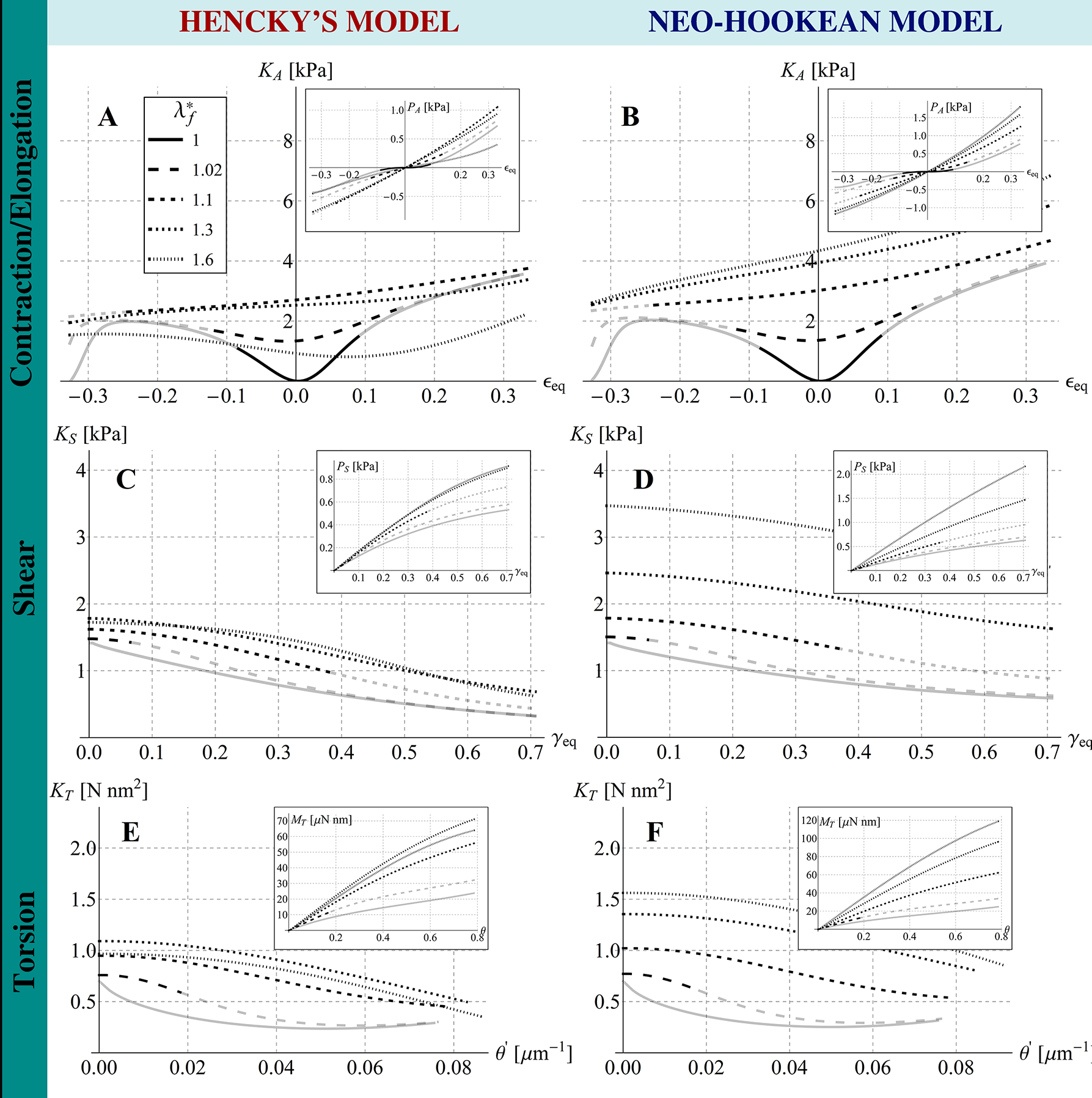}} 
\caption{\textbf{A-B)} Equivalent axial stiffness $K_A$ (and, in the inset, nominal stress $P_A$) against equivalent strain $\varepsilon_{eq}$, in case of  contraction/elongation and for both Hencky (H) and neo-Hookean (NH) models. \textbf{C-D)} Equivalent shear modulus $K_S$ (and, in the inset, nominal stress $P_S$) versus the equivalent shear strain $\gamma_{eq}$, for Hencky and neo-Hookean elements. \textbf{E-F)} Torsional stiffness and twisting moment as functions of the unit torsion angle $\theta'$ and torsion angle $\theta$, respectively, for the two hyperelastic laws hypothesized. The analyses were performed for different values of cables pre-stretch $\lambda^*_f\,(1, 1.02, 1.1, 1.3, 1.6)$, by making reference to the cell parameters collected in Table \ref{tab.Phisical_parameters}. In lighter gray the tracts of the curves theoretically extrapolated but unrealistic since therein cables would undergo compression in a number such as to impair the equlibrium of the whole system.}
\label{fig.K} 
\end{figure}

\begin{figure} [htbp]
\centering
{\includegraphics[width=1\textwidth]{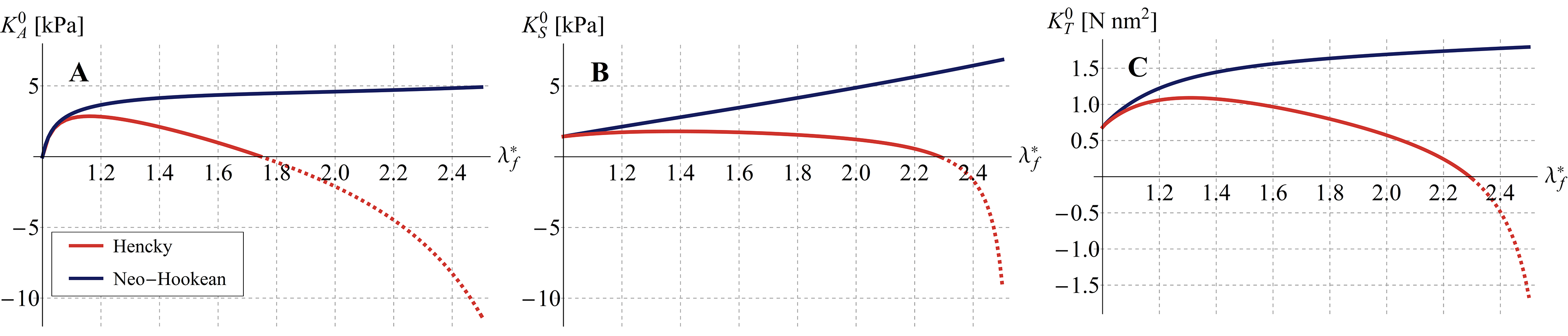}} 
\caption{Tangent (initial) stiffness exhibited by the soft-strut $30$-element tensegrity by varying the cables pre-stretch $\lambda^*_f$, for Hencky (H) and neo-Hookean (NH)  models:\textbf{A)} contraction/elongation, \textbf{B)} shear and \textbf{C)} torsion. The analyses were performed for values of the parameters as reported in Table \ref{tab.Phisical_parameters}. The dashed parts of the curves relative to the Hencky's model highlight negative stiffness values.}
\label{fig.K0} 
\end{figure}

By considering that the cellular tensegrity model is virtually tested to torque, it is twisted by prescribing a growing torsion angle $\theta$ at the top of the structure through proper displacements imposed at the upper nodes $1, 5$ and $9$, keeping the nodes at the basis locked. In this case, to obtain the cell response, we conveniently start from the updated nodes' coordinates, given by 
\begin{equation}
\textbf{p}_i^U=\left( R_i cos\left(\dfrac{\pi}{2} - \alpha_i - \Delta\alpha_i\right), R_i sin\left(\dfrac{\pi}{2} - \alpha_i - \Delta\alpha_i\right), z_i + w_i \right),
\end{equation}
where $R_i$ represents the radius of the circle passing through the nodes lying in the $\{x_R,y_R\}$ plane, in particular being $R_i=R_T$ for the nodes belonging to the upper and lower equilateral triangles, $R_i=R_{HT}$ for the nodes $3,7,11$ of the middle hexagon and $R_i=R_{HB}$ for the nodes $2,6,10$. Also, $\alpha_i$ is the angle that the generic node $i$ forms with respect to the $y_R$ axis, $\Delta\alpha_i$ describing the corresponding incremental angle (clockwise, whence the minus) due to the torsional rotation. Additionally, it is possible to assume that, for the nodes placed at the same height, the vertical displacements $w_i$ are the same, so that also in this case the relations \eqref{vertical_disp} and the \eqref{vert_disp2}$_{3}$ hold true. Moreover, geometrical arguments allow to set:
\begin{align}
\label{dn_dr_torsione}
\notag &\delta_{HT} = \Delta\alpha_3 = \Delta\alpha_7 = \Delta\alpha_{11}, \\
&\delta_{HB} = \Delta\alpha_2 = \Delta\alpha_6 = \Delta\alpha_{10}. 
\end{align}
Because the basis of the structure is constrained, the unknowns of the problem finally reduce to the vertical displacements $ w_{UT} $, $ w_{HT} $ and $ w_{HB} $, the torsion angles $\delta_{HT}$ and $\delta_{HB}$ and the radii $R_{HT}$ and $R_{HB}$.\\
The results are shown in Figure \ref{fig.DefPlot} in terms of overall deformation, for $\lambda_f^* = 1.1$ and a prescribed torsion angle $\theta$ which varies up to $\pi/4$.
The torsional stiffness $ K_T $, computed as first derivative of the twisting moment $ M_T $ with respect to the unit torsion angle $ \theta' = \theta/h $, is shown in Figure \ref{fig.K}\hyperref[fig.K]{E-F} for different values of cables pre-stretch and for both Hencky and neo-Hookean laws. The corresponding twisting moment of the cell structure --obtained as $M_T := \partial U/\partial \theta$-- is plotted against $ \theta$ in the insets. Similarly to the case of shear, a decrease of the torsional stiffness as the rotation increases is observed. In particular, it is still found that the tensegrity system whose elements obey the Hencky model exhibits a lowering of its initial (tangent) torsional stiffness at large pre-stretches, the neo-Hookean tensegrity instead showing a significant stiffness increase for the same pre-stretch values, as shown in Figure \ref{fig.K0}\hyperref[fig.K0]{C}. As for the shear, we also highlight that, if the system is initially characterized by an unit pre-stretch, torque would induce compression at the early stage of the prescribed torsional rotation in a number of cables such that the whole structure would no longer be able to guarantee equilibrium, with the result that, in absence of pre-stress, the initial (tangent) torsional stiffness $K_T^0$ must be vanishing.

\section{Symmetry losing: local buckling and global configurational switching}

\begin{figure} [htbp]
\centering
{\includegraphics[width=1\textwidth]{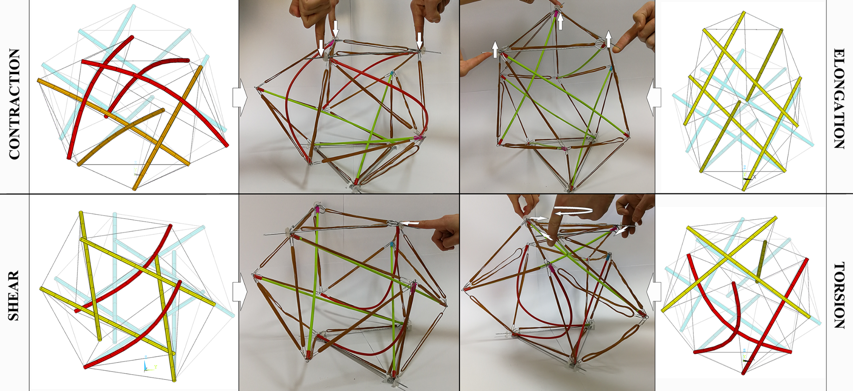}} 
\caption{Experimental responses of handmade toy tensegrity systems with bendable bars experiencing elongation, contraction, shear and torsion highlighting possible local buckling and global loss of symmetry, with comparison of deformations obtained from FE analyses.}
\label{fig.toysystem} 
\end{figure}

So far we have analyzed how cell cytoskeleton would behave by expecting that its pre-stressed structure, modeled as a soft-strut tensegrity system, preserves symmetries in both self-equilibrated (form-finding) states and undergoing deformations in response to applied loads. This implied that bending of microtubules under compression was neglected, being enabled their sole elastic shortening. Also, equilibria associated to global switching of the tensegrity on possible not symmetrical configurations minimizing the elastic energy were not explored.\\
Local loss of symmetry is however not an unrealistic event in cells.
In fact, according to Table \ref{tab.Phisical_parameters} and experimental measures \cite{Gittes_1993}, we considered the effective geometry of the cross section of the microtubules as possessing a length $L_t$ and a bending stiffness $B_t=2.15 \, 10^{-23} \, N \cdot m^2$, so obtaining a critical axial load due to instability as $N_c \simeq 1.5 \, pN$. Such a value is compatible with the order of magnitude of the forces occurring in the struts both when the tensegrity is at self-equilibrium and when it is solicited by external loads, this legitimating the possibility that a post-buckling response cooperates with the purely axial contraction of the bars in influencing the actual cell mechanical behavior. 
As a matter of fact, buckling of cell microtubules has been observed experimentally \cite{Wang2001MT,Stamenovic_2001} and theoretically investigated in some literature works \cite{Coughlin1997,Coughlin1998,Volokh2000}, by however postulating the axial rigidity of the struts. 
More recent studies \cite{Brangwynne_2006,Li_2008} highlighted that the critical load induced by buckling of microtubules embedded in the cellular environment would turn out to be significantly greater (from about two up to four orders of magnitude) than the one evaluated for the same isolated element. This difference would occur due to the presence of the surrounding viscous/viscoelastic cytoplasm, which also comprises the elastic network of intermediate filaments. These, together with other intracellular proteic structures of the gel-like cytosol, would therefore work as a tensed lateral support that stabilizes microtubules, so increasing their effective capability to resist buckling \cite{Brodland1990, Stamenovic_2001, Brangwynne_2006, Li_2008} and in some cases forcing the microtubules to not buckle in a single-wave mode\cite{Soheilypour_2015}.
Local buckling of struts is however not the sole way for envisaging loss of symmetry in a cellular tensegrity structure. Equilibria could in fact be reached --at least in principle-- during any deformation process in cells when prescribed levels of pre-stress and tensile forces, respectively in bars and cables, attain values such that the tensegrity is invited to deviate from its natural shape to follow minimal energy pathways. This is for instance the case of experimentally observed overall configurational switching of cells occurring during gastrulation \cite{Wolgemuth2011} or in adhesion and migration phenomena, in occasion of which abrupt changes of cytoskeleton organization are required to accommodate polymerization/depolymerization of protein filaments to respond to specific chemo-mechanical stimuli resulting in reorientation of the stress fibers \cite{plosone,KAUNAS2009,Qian_2013}.\\
With all this in mind, in order to explore both local and global loss of symmetry situations in soft-strut cellular tensegrity systems, we therefore analyze two hypothetical scenarios, that \textit{de facto} we will find can occur both separately and as concomitant: the case of soft struts experiencing buckling and the case of minimum energy equilibrium states associated to overall deviation of the tensegrity from its expected (symmetrical) configuration. Our theoretical results prove that the interplay of axial deformability of soft bars and bending stiffness might actually trigger complex behaviors and allow not symmetrical cytoskeleton shapes, guided by the competition among local and global instability phenomena, as also roughly confirmed by the responses of the handmade toy system in Figure \ref{fig.toysystem}.\\

\subsection{Competition of local buckling and not symmetrical shapes in cellular tensegrity systems with bendable soft struts: form-finding and response to applied loads}

\begin{figure} [htbp]
\centering
{\includegraphics[width=1\textwidth]{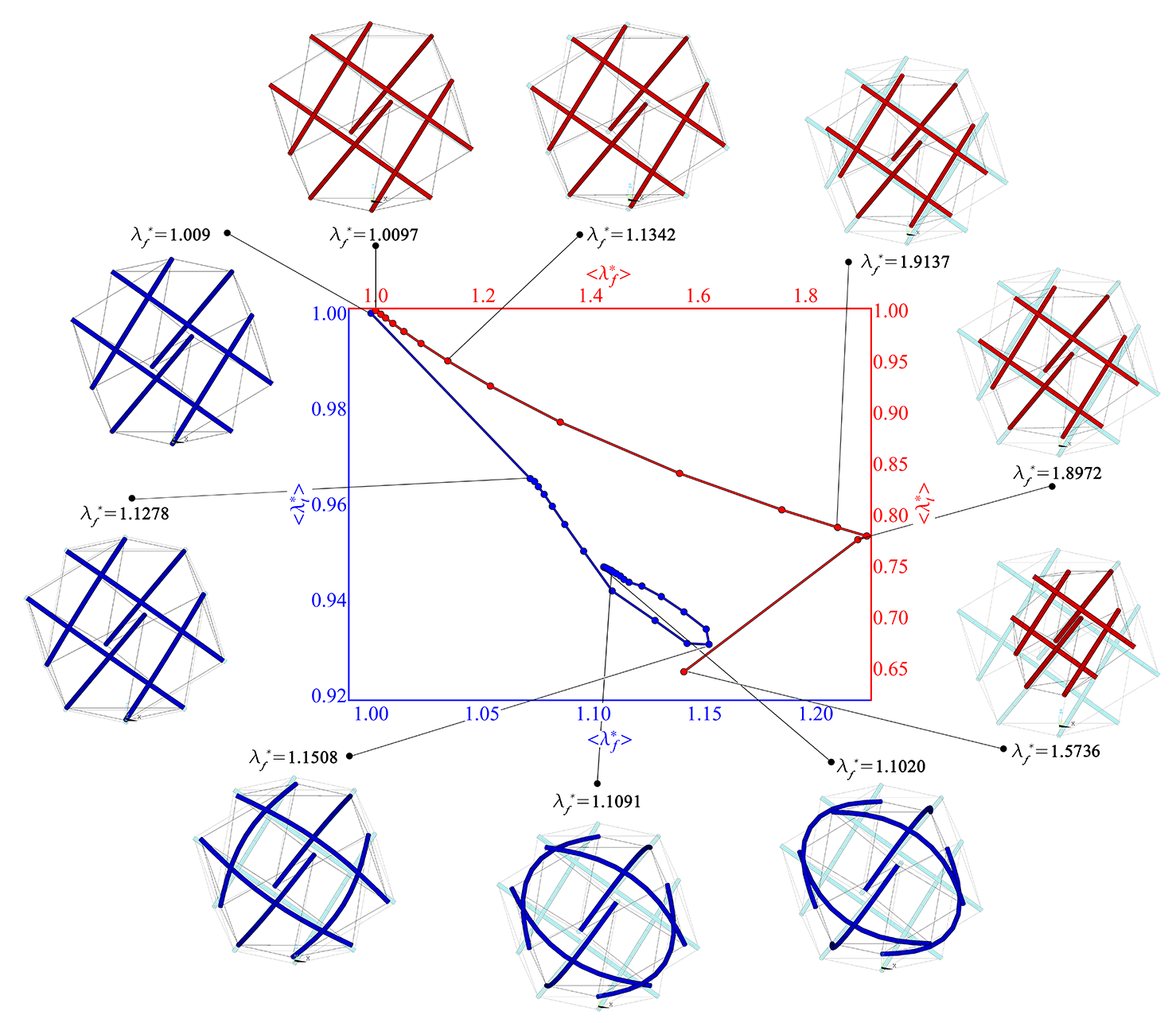}} 
\caption{Form-finding and equilibria of the soft-strut cellular tensegrity represented in the phase space $\langle\lambda_f^*\rangle - \langle\lambda_t^*\rangle$. Blue points and (fitting) curves represent the states of self-equilibrium that the system gains for each pair of (average) elongation and contraction pre-stretches in its cables and bars, respectively, for the case in which the bending stiffness of the struts is assumed up to values of about $10^4$ times \cite{Brangwynne_2006, Li_2008} the geometrical bending stiffness $B_t$ experimentally measured for an isolated microtubule \cite{Gittes_1993}. Red points and associated fitting curves represent instead the self-equilibrium states in case the tensegrity struts can axially deform but their bending stiffness $B_t \rightarrow \infty$. Note that in both the cases form-finding provides possible loss of symmetry: this happens, at the local level, with buckling of microtubules (deformed structures with blue struts) for bendable bars and with global configurational switching (tensegrities with red struts) if the bending stiffness of microtubules is forced to be extremely high. To synoptically show the two behaviors in the same phase space, two corresponding different scales and colors (blu and red) are utilized for the axes.}
\label{fig.phasespace} 
\end{figure}

We start by reanalyzing the form-finding problem of a soft-bar tensegrity structure whose contracting struts are now enabled to also undergo bending.
To make this coherently with experimental data and observations and to properly take into account the effect of the lateral confinement imposed by cytoplasm and other proteic structures to microtubules of actual living cells, we ad hoc considered a fictitious amplification of the geometrical bending stiffness $B_t$ up to $10^4$ \cite{Brangwynne_2006, Li_2008}, leaving unchanged the cross-sectional areas of the microtubules, responsible of their axial deformability. Higher values of the \textit{effective} bending stiffness of the microtubules were then additionally assumed, in this manner allowing the cell cytoskeleton, even though the axially soft struts were not prone to bend \cite{Soheilypour_2015}, to homotetically scale its polyhedral shape as a function of the increasing pre-stretch in the cables, or to switch asymmetrically on other energetically more comfortable configurations.\\
Symmetry losing no longer allowed to proceed analytically and therefore we made reference to Finite Element numerical simulations. All the analyses were therefore performed by reconstructing the three-dimensional icosahedral $30$-element tensegrity structure with the aid of the Finite Element commercial code ANSYS\textsuperscript{\textregistered} \cite{ANSYS}, uploading a progressively growing level of elastic pre-stretch to the tensed filaments (cables) and so inducing a corresponding increasing compressive pre-stress in the microtubules (struts), that --from the operational point of view-- was managed by properly tuning the initial (at rest) cables lengths on the basis of the geometrical relations involving stretches and resting lengths already established above. Non-linearly elastic elements bar (LINK180) with no-compression and axially deformable and bending beams (BEAM188) were hence chosen to replicate respectively filaments and microtubules, using for both the Hencky's hyperelastic law \cite{Plesek2006} and assigning to cables and struts the corresponding geometrical features. At the end, a mesh resulting in $84$ elements and $66$ nodes with translational and rotational degrees of freedom was generated. All the numerical analyses were conducted in finite strains and large displacements, by activating the options of nonlinear geometry, standard step-by-step procedures and robust algorithms furnished by the software to control and ensure the convergence. Moreover, a preliminary check was performed to verify that the model was capable to confirm the theoretical results, utilized as benchmark, already obtained for the case of symmetry-preserving deformations.\\
The results are synoptically shown in Figure \ref{fig.phasespace} in which all possible self-equilibrated states that the tensegrity system can assume are uniquely represented by points in the phase space $\langle\lambda_f^*\rangle - \langle\lambda_t^*\rangle$, where $\langle\lambda_f^*\rangle$ and $\langle\lambda_t^*\rangle$ are the pre-stretch average values in filaments and tubules, respectively.\\

In this phase space, by starting from slightly higher-than-one levels of average pre-stretch in the filaments, the corresponding average contractions almost proportionally grow in microtubules, at the early stage of the pre-stress showing that the soft-strut tensegrity overall contracts homotetically, preserving shape and the original polyhedral symmetry. This behavior is exhibited by the system up to prestretches in filaments $\lambda_f^*\simeq 1.13$, for both the cases of bendable and stiff struts, that for the effective bending stiffness four orders of magnitude greater than the geometrical one $B_t=2.15 \, 10^{-23} \, N \times m^2$, and for ideally unbendable struts, say $B_t\rightarrow\infty$. However, as the pre-stretch in the filaments increases, very different behaviors are exhibited by the system in the two cases of bendable and unbendable microtubules. In fact, in the first case, as filaments pre-stretch grows, bars elastically contract by increasing the compressive stress they sustain, then suddenly undergo buckling preserving part of the axial contraction and producing a sharp snap-back phenomenon at $\lambda_f^*\simeq 1.15$, at the end progressively relaxing the axial deformation (see blue curve in Figure \ref{fig.phasespace}). However, the local buckling of the struts in this case occurs for all the compressed elements contemporaneously and this allows the tensegrity to maintain its overall symmetrical shape. In the other case, say when the microtubules nominal bending stiffness is set to be high, the early stage of the deformation is still characterized by simple uniform scaling of the polyhedral tensegrity shape, up to a pre-stretch value in the filaments of about $\lambda_f^*\simeq 1.9$, after which an abrupt change of configuration is exhibited by the structure which switches on a deformed state associated with loss of symmetry (see red curve in Figure \ref{fig.phasespace}), then finding a stable equilibrium after a reversal in the phase space, by leaping up lower pre-stretch levels and higher contraction of microtubules.\\

\begin{figure} [htbp]
\centering
{\includegraphics[width=1\textwidth]{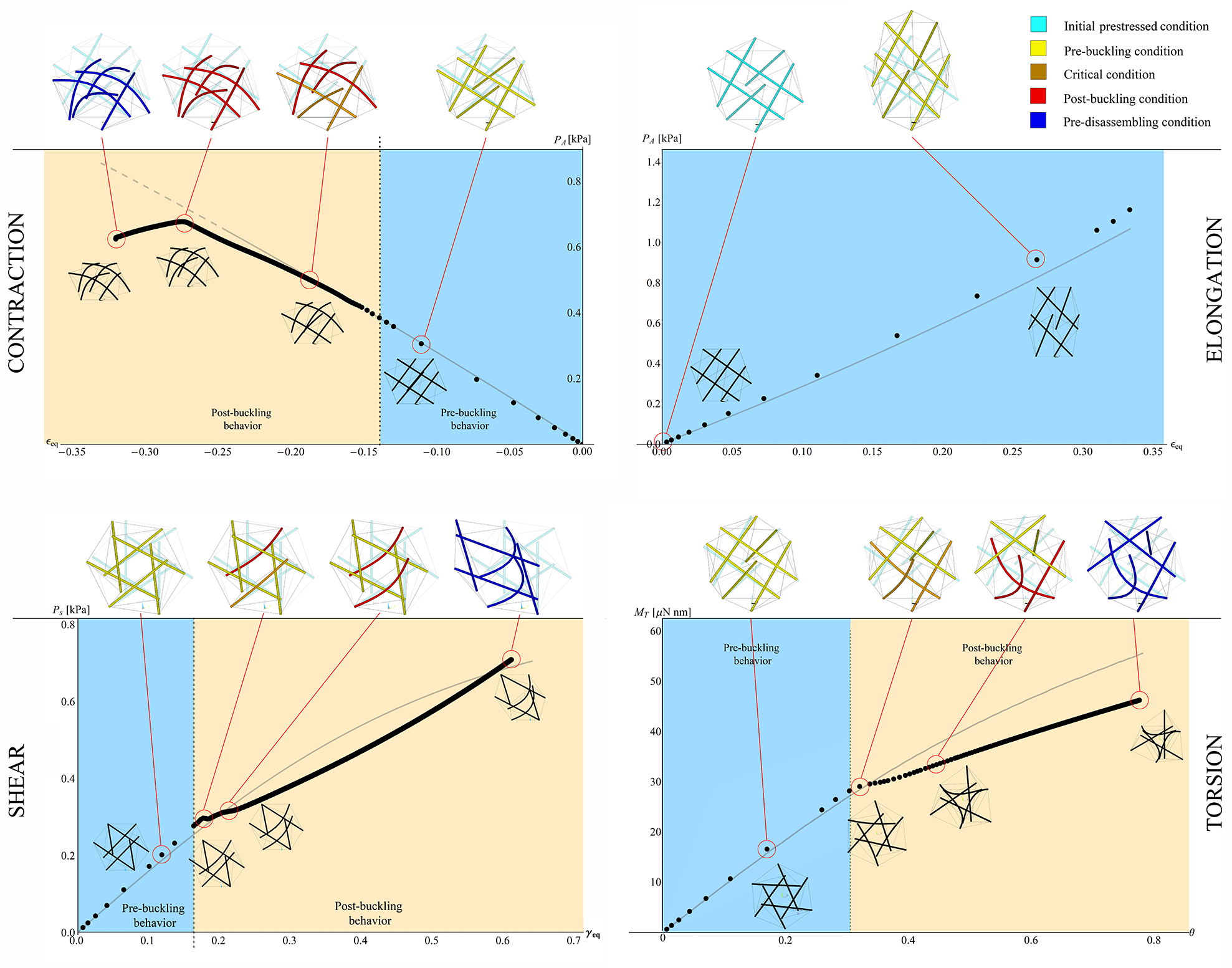}} 
\caption{Numerical FE-based (black dots) results for soft tensegrity systems with bendable struts in cases of contraction, elongation, shear and torsion, in terms of generalized stresses (applied axial forces, shear loads and torque) versus corresponding equivalent strains (overall contraction/elongation axial strains, shear deformation and global torsion angle). Grey curves recall the solution in the cases of preservation of the expected symmetries, as obtained from the theoretical analyses in absence of buckling struts. Blue and flesh-colored backgrounds allow to distinguish the so-called pre-buckling and post-buckling regions, respectively. The insets and the three-dimensional sketches on the top of each graphic show how tensegrities behave, in terms of deformation, as the loads increase.}
\label{fig.not_sym_results} 
\end{figure}

To finally explore what happens if the bendability of the struts is taken into account also for soft tensegrity structures undergoing the same applied forces already considered above in case of (imposed) symmetry-preserved situations, we numerically replicated the analyses for the entire set of loading conditions, say contraction, elongation, shear and torsion, by performing the simulations via FEM.\\
The results, in terms of overall deformation of the system, are illustrated for each of the four load cases in Figure \ref{fig.toysystem}, to highlight the qualitative compatibility of the obtained deformed configurations with respect to those shown by the handmade toy systems roughly loaded with the corresponding forces. More in details, quantitative results are reported --in terms of generalized stresses against associated overall strains-- in Figure \ref{fig.not_sym_results}. It is therein worth noticing that, as already found for the form-finding problem, we can observe a first phase of the mechanical response where an essentially perfect superposition of the numerical FE results with the outcomes obtained theoretically is traced until the pre-stress in cables and struts is such that the minimization of the total potential energy of the structure can be still attained for symmetry-preserving configurations. Then, a second phase can be registered in which the tensegrity system progressively undergoes no longer symmetrical deformation states as the applied loads increase and induce both buckling of some strut elements and global switching of the whole structures. In particular, with reference to the overall deformation ranges considered in the analyses, except for the case of elongation, in all the other loading conditions we can sharply separate a \textit{pre-buckling} phase, recopying the already obtained analytical results related to symmetry-preserving equilibrium states (highlighted by blue background in the graphics), from a \textit{post-buckling} behavior (denoted by the flesh-colored background), characterized by a deviation of the points (black dots) from the curve (grey line) which denotes the path ideally followed by the structure in case of absence of buckling of struts. The results are truncated at an end point in correspondence of which cables and struts are no longer able to sustain stresses for ensuring global equilibrium in the actual (deformed) configuration. To add geometrical information about what physically happens during the load increase, in Figure \ref{fig.not_sym_results} insets with plane and three-dimensional views of the tensegrity systems, at any relevant stage of the deformation, are supplied.

\section{Conclusions}
The cytoskeleton is a complex, continuously reorganizing and self-assembling network of interconnected microtubules, actin filaments and microfilaments to which is assigned --among other-- the role of bearing structure of living cells. In response to chemo-mechanical stimuli, the cytoskeletal elements activate polymerization/depolymerization as well as micro-structural disarrangements \cite{prsa2018} processes and undergo large deformations accompanied by elastic pre-stretch to form either stable configurations to govern adhesion and ensure equilibrium of internal stresses and applied forces coming from cell-ECM and cell-cell interactions, or unstable shapes, for driving migration phenomena, cell reorientation and duplication phenomena, at any of these stages the cytoskeleton architecture contemporary providing energy storing and allowing selected signaling pathways across the cell membrane and towards the nucleus.\\
By starting from the Ingber's idea of using tensegrity systems for describing the mechanical behavior of the cell cytoskeleton and studying how equilibria evolve as its structural geometry changes, we built up a new \textit{soft}-strut tensegrity model of the cell cytoskeleton. With the aim to overcome some limits of previous models related to intrinsic (constitutive and kinematical) assumptions, we in particular removed the hypothesis of linear elasticity for both cables and bars, rewriting the form-finding problem as well as the equations governing the elastic response to applied loads of a $30$-element polyhedral tensegrity structure by including both axial deformability and bendability of struts, coherently with experimental measures --that highlighted close values of axial stiffness for single actin filaments and microtubules-- and according to recent literature findings showing \textit{in vivo} buckling of microtubules.\\ \\
The analytical and numerical results provided by the soft tensegrity paradigm of the cell, as in detail described in the previous Sections, showed that rich families of nonlinear elastic responses of the cell cytoskeleton, presenting some previously unpredicted non-monotonic overall stress-strain curves and softening phenomena, can be derived by following its rearrangement under external actions.  Also, FE analyses demonstrated that the loss of both local and global symmetry of the tensegrity structure can be found as a consequence of non-uniform buckling of its elements and configurational switching of the whole system on asymmetric shapes. All these transitions occur in correspondence of energy wells generated by a complex competition among bending stiffness and axial deformability of the struts, average pre-stress levels in filaments and microtubules and structure instabilities.
However, in the present paper, the Authors did not take into account additional aspects that would certainly enrich the modeling of the mechanical response of the tensegrity-based cytoskeleton. In fact, more complex tensegrity-based architectures should be incorporated for properly and faithfully modeling the mechanical behavior of the cell cytoskeleton and coherently catching the dynamics characterizing the actual internal reorganization and distribution of forces during the cell dynamics. We did not consider the effects of the intrinsic viscoelasticity of the tensegrity elements, the dissipation being potentially governed by the presence of the viscous cytosol. Also, the entropic elasticity of the filaments --related to the folding and unfolding of their polymeric chains-- has not been introduced, a fact that could be explicitly modeled by providing a biochemical-mechanical coupling, by means of which chemical gradients can be used to drive the polymerization/depolymerization of the cytoskeletal filaments in different cell districts.
It is felt that the proposed enhanced tensegrity model, by allowing to quantitatively predict the order of magnitude of forces, stiffness and elastic energy amount stored by the pre-stressed cell cytoskeleton and being also capable to replicate both symmetry-preserving and instability-guided asymmetric configurations of the protein structural network, could contribute to move a further step towards an engineering modeling of adhesion and migration of single cells and shed light on the underlying physics of many important phenomena not yet fully understood, involving abrupt changes of cytoskeleton configurations or cell morphology, such as gastrulations, extreme deformations occurring during duplication and modifications of elastic properties characterizing physiological cell processes and malignant transformations of cancer and metastatic cells \cite{FRALDI20181,CAROTENUTO20181}.

\paragraph*{Acknowledgments} 
MF gratefully acknowledges the support of the grants by MIUR ARS01-01384-PROSCAN and by the University of Napoli "Federico II" E62F17000200001-NAPARIS. LD gratefully acknowledges the support of the grant ERC-2013-ADG-340561-INSTABILITIES. NP gratefully acknowledges the support of the grants by the European Commission Graphene Flagship Core 2 n. 785219 (WP14 "Composites") and FET Proactive "Neurofibres" n. 732344 as well as of the grant by MIUR "Departments of Excellence" grant L. 232/2016.

\newpage
\bibliographystyle{unsrt}  

\begin{thebibliography}{10}

\bibitem{mechbook}
M.~R.~K. Mofrad and R.~D. Kamm, eds., {\em Cellular Mechanotransduction}.
\newblock Cambridge University Press ({CUP}), 2009.

\bibitem{C3SM52769G}
R.~H. Pritchard, Y.~Y. Shery~Huang, and E.~M. Terentjev, ``Mechanics of
  biological networks: from the cell cytoskeleton to connective tissue,'' {\em
  Soft Matter}, vol.~10, pp.~1864--1884, 2014.

\bibitem{Fraldi_2015}
M.~Fraldi, A.~Cugno, L.~Deseri, K.~Dayal, and N.~M. Pugno, ``A frequency-based
  hypothesis for mechanically targeting and selectively attacking cancer
  cells,'' {\em J. R. Soc. Interface}, vol.~12, p.~20150656, sep 2015.

\bibitem{Fraldi_2017}
M.~Fraldi, A.~Cugno, A.~R. Carotenuto, A.~Cutolo, N.~M. Pugno, and L.~Deseri,
  ``Small-on-large fractional derivative{\textendash}based single-cell model
  incorporating cytoskeleton prestretch,'' {\em Journal of Engineering
  Mechanics}, vol.~143, p.~D4016009, may 2017.

\bibitem{Stamenovic}
D.~Stamenovi{\'{c}}, ``Models of cytoskeletal mechanics based on tensegrity,''
  in {\em Cytoskeletal Mechanics} (M.~R.~K. Mofrad and R.~D. Kamm, eds.),
  pp.~103--128, Cambridge University Press ({CUP}), 2006.

\bibitem{nih}
D.~E. Ingber, N.~Wang, and D.~Stamenovic, ``{Tensegrity, cellular biophysics,
  and the mechanics of living systems.},'' {\em Reports on progress in physics.
  Physical Society (Great Britain)}, vol.~77, no.~4, p.~046603, 2014.

\bibitem{mofrad2006cytoskeletal}
M.~R. Mofrad and R.~D. Kamm, {\em Cytoskeletal mechanics: models and
  measurements in cell mechanics}.
\newblock Cambridge University Press, 2006.

\bibitem{Ingber_2003}
D.~E. Ingber, ``Tensegrity {I}. cell structure and hierarchical systems
  biology,'' {\em Journal of Cell Science}, vol.~116, pp.~1157--1173, apr 2003.

\bibitem{Wang_2009}
N.~Wang, J.~D. Tytell, and D.~E. Ingber, ``Mechanotransduction at a distance:
  mechanically coupling the extracellular matrix with the nucleus,'' {\em
  Nature Reviews Molecular Cell Biology}, vol.~10, pp.~75--82, jan 2009.

\bibitem{Wang2001MT}
N.~Wang, K.~Naruse, D.~Stamenovic, J.~J. Fredberg, S.~M. Mijailovich, I.~M.
  Tolic-Norrelykke, T.~Polte, R.~Mannix, and D.~E. Ingber, ``Mechanical
  behavior in living cells consistent with the tensegrity model,'' {\em
  Proceedings of the National Academy of Sciences}, vol.~98, pp.~7765--7770,
  jul 2001.

\bibitem{Tee_2015}
Y.~H. Tee, T.~Shemesh, V.~Thiagarajan, R.~F. Hariadi, K.~L. Anderson, C.~Page,
  N.~Volkmann, D.~Hanein, S.~Sivaramakrishnan, M.~M. Kozlov, and A.~D.
  Bershadsky, ``Cellular chirality arising from the self-organization of the
  actin cytoskeleton,'' {\em Nature Cell Biology}, vol.~17, pp.~445--457, mar
  2015.

\bibitem{Xu_2016}
G.-K. Xu, B.~Li, X.-Q. Feng, and H.~Gao, ``A tensegrity model of cell
  reorientation on cyclically stretched substrates,'' {\em Biophysical
  Journal}, vol.~111, pp.~1478--1486, oct 2016.

\bibitem{Xu_2018}
G.-K. Xu, X.-Q. Feng, and H.~Gao, ``Orientations of cells on compliant
  substrates under biaxial stretches: A theoretical study,'' {\em Biophysical
  Journal}, vol.~114, pp.~701--710, feb 2018.

\bibitem{Ingber_1981}
D.~E. Ingber, J.~A. Madri, and J.~D. Jamieson, ``Role of basal lamina in
  neoplastic disorganization of tissue architecture.,'' {\em Proceedings of the
  National Academy of Sciences}, vol.~78, pp.~3901--3905, jun 1981.

\bibitem{Ingber_1985}
D.~E. Ingber and J.~D. Jamieson, ``Cells as tensegrity structures:
  Architectural regulation of histodifferentiation by physical forces
  transduced over basement membrane.,'' {\em Gene Expression during Normal and
  Malignant Differentiation.}, 1985.

\bibitem{Ingber_1993b}
D.~E. Ingber, ``Cellular tensegrity: defining new rules of biological design
  that govern the cytoskeleton,'' {\em Journal of Cell Science}, 1993.

\bibitem{Gray}
J.~Gray, {\em How {A}nimals {M}ove}.
\newblock Cambridge University Press, 1953.

\bibitem{FRALDI20131310}
M.~Fraldi, F.~Carannante, and L.~Nunziante, ``Analytical solutions for n-phase
  functionally graded material cylinders under de saint venant load conditions:
  Homogenization and effects of poisson ratios on the overall stiffness,'' {\em
  Composites Part B: Engineering}, vol.~45, no.~1, pp.~1310 -- 1324, 2013.

\bibitem{motro}
R.~Motro, {\em Tensegrity: structural systems for the future}.
\newblock Kogan Page Science, 2003.

\bibitem{skelton}
R.~E. Skelton and M.~C. Oliveira, {\em Tensegrity Systems}.
\newblock Springer Nature, 2009.

\bibitem{Fraternali2015}
F.~Fraternali, G.~Carpentieri, and A.~Amendola, ``{On the mechanical modeling
  of the extreme softening/stiffening response of axially loaded tensegrity
  prisms},'' {\em Journal of the Mechanics and Physics of Solids}, vol.~74,
  pp.~136--157, 2015.

\bibitem{Liu_2017}
K.~Liu, J.~Wu, G.~H. Paulino, and H.~J. Qi, ``Programmable deployment of
  tensegrity structures by stimulus-responsive polymers,'' {\em Scientific
  Reports}, vol.~7, jun 2017.

\bibitem{Stamenovic1996}
D.~Stamenovi{\'{c}}, J.~J. Fredberg, N.~Wang, J.~P. Butler, and D.~E. Ingber,
  ``A microstructural approach to cytoskeletal mechanics based on tensegrity,''
  {\em Journal of Theoretical Biology}, vol.~181, pp.~125--136, jul 1996.

\bibitem{Luo2008}
Y.~Luo, X.~Xu, T.~Lele, S.~Kumar, and D.~E. Ingber, ``A multi-modular
  tensegrity model of an actin stress fiber,'' {\em Journal of Biomechanics},
  vol.~41, pp.~2379--2387, aug 2008.

\bibitem{violinbow}
S.~Palumbo, A.~R. Carotenuto, A.~Cutolo, L.~Deseri, and M.~Fraldi, ``Nonlinear
  elasticity and buckling in the simplest soft-strut tensegrity paradigm,''
  {\em International Journal of Non-Linear Mechanics}, 2018.

\bibitem{Coughlin1997}
M.~F. Coughlin and D.~Stamenovic, ``A tensegrity structure with buckling
  compression elements: Application to cell mechanics,'' {\em Journal of
  Applied Mechanics}, vol.~64, no.~3, p.~480, 1997.

\bibitem{Coughlin1998}
M.~F. Coughlin and D.~Stamenovic, ``A tensegrity model of the cytoskeleton in
  spread and round cells,'' {\em Journal of Biomechanical Engineering},
  vol.~120, no.~6, p.~770, 1998.

\bibitem{Stamenovic2000}
D.~Stamenovic and M.~F. Coughlin, ``A quantitative model of cellular elasticity
  based on tensegrity,'' {\em Journal of Biomechanical Engineering}, vol.~122,
  no.~1, p.~39, 2000.

\bibitem{Volokh2000}
K.~Volokh, O.~Vilnay, and M.~Belsky, ``Tensegrity architecture explains linear
  stiffening and predicts softening of living cells,'' {\em Journal of
  Biomechanics}, vol.~33, pp.~1543--1549, dec 2000.

\bibitem{Stamenovic_2001}
D.~Stamenovic, S.~M. Mijailovich, I.~M. Tolic-Norrelykke, J.~Chen, and N.~Wang,
  ``Cell prestress. {II}. contribution of microtubules,'' {\em {AJP}: Cell
  Physiology}, vol.~282, pp.~C617--C624, oct 2001.

\bibitem{Liu2002}
X.~Liu and G.~H. Pollack, ``Mechanics of f-actin characterized with
  microfabricated cantilevers,'' {\em Biophysical Journal}, vol.~83,
  pp.~2705--2715, nov 2002.

\bibitem{Deguchi2006}
S.~Deguchi, T.~Ohashi, and M.~Sato, ``Tensile properties of single stress
  fibers isolated from cultured vascular smooth muscle cells,'' {\em Journal of
  Biomechanics}, vol.~39, pp.~2603--2610, jan 2006.

\bibitem{Brodland1990}
G.~Brodland and R.~Gordon, ``Intermediate filaments may prevent buckling of
  compressively loaded microtubules,'' {\em Journal of Biomechanical
  Engineering}, vol.~112, p.~319, aug 1990.

\bibitem{holzapfel}
G.~A. Holzapfel, {\em Nonlinear Solid Mechanics: A Continuum Approach for
  Engineering}.
\newblock Wiley, 2000.

\bibitem{Gardel2008}
M.~L. Gardel, K.~E. Kasza, C.~P. Brangwynne, J.~Liu, and D.~A. Weitz,
  ``{Mechanical Response of Cytoskeletal Networks},'' {\em Methods in Cell
  Biology}, vol.~89, no.~08, pp.~487--519, 2008.

\bibitem{Hill_1968}
R.~Hill, ``On constitutive inequalities for simple materials-{I},'' {\em
  Journal of the Mechanics and Physics of Solids}, vol.~16, pp.~229--242, aug
  1968.

\bibitem{Bigoni2012}
D.~Bigoni, {\em {Nonlinear Solid Mechanics - Bifurcation theory and material
  instability}}.
\newblock 2012.

\bibitem{Plesek2006}
J.~Ple{\v{s}}ek and A.~Kruisov{\'{a}}, ``Formulation, validation and numerical
  procedures for {H}encky's elasticity model,'' {\em Computers {\&}
  Structures}, vol.~84, pp.~1141--1150, jun 2006.

\bibitem{Xiao}
A.~M. O.~T.~Bruhns, H.~Xiao, ``Constitutive inequalities for an isotropic
  elastic strain-energy function based on hencky's logarithmic strain tensor,''
  {\em Proceedings of the Royal Society A: Mathematical, Physical and
  Engineering Sciences}, 2001.

\bibitem{Anand1979}
L.~Anand, ``On {H}. {H}encky's approximate strain-energy function for moderate
  deformations,'' {\em Journal of Applied Mechanics}, vol.~46, no.~1, p.~78,
  1979.

\bibitem{Gittes_1993}
F.~Gittes, ``Flexural rigidity of microtubules and actin filaments measured
  from thermal fluctuations in shape,'' {\em The Journal of Cell Biology},
  vol.~120, pp.~923--934, feb 1993.

\bibitem{Rolfe_1997}
D.~F. Rolfe and G.~C. Brown, ``Cellular energy utilization and molecular origin
  of standard metabolic rate in mammals,'' {\em Physiological Reviews},
  vol.~77, pp.~731--758, jul 1997.

\bibitem{Bernstein_2003}
B.~W. Bernstein and J.~R. Bamburg, ``Actin-{ATP} hydrolysis is a major energy
  drain for neurons,'' {\em The Journal of Neuroscience}, vol.~23, pp.~1.2--6,
  jan 2003.

\bibitem{natgeo}
M.~Greshko, ``How many cells are in the human body—and how many microbes?.''
  \url{https://news.nationalgeographic.com/2016/01/160111-microbiome-estimate-count-ratio-human-health-science/},
  January 2016.

\bibitem{energy}
C.~D. A., B.~W. W., R.~C. L., A.~Jordi, and R.~K. J., {\em Mass Transport:
  Circulatory System with Emphasis on Nonendothermic Species}, pp.~17--66.
\newblock American Cancer Society, 2016.

\bibitem{Hooper_2008}
S.~L. Hooper, K.~H. Hobbs, and J.~B. Thuma, ``Invertebrate muscles: Thin and
  thick filament structure, molecular basis of contraction and its regulation,
  catch and asynchronous muscle,'' {\em Progress in Neurobiology}, vol.~86,
  pp.~72--127, oct 2008.

\bibitem{Pollard_2003}
T.~D. Pollard and G.~G. Borisy, ``Cellular motility driven by assembly and
  disassembly of actin filaments,'' {\em Cell}, vol.~113, p.~549, may 2003.

\bibitem{Abraham_1999}
V.~C. Abraham, V.~Krishnamurthi, D.~L. Taylor, and F.~Lanni, ``The actin-based
  nanomachine at the leading edge of migrating cells,'' {\em Biophysical
  Journal}, vol.~77, pp.~1721--1732, sep 1999.

\bibitem{Svitkina_1997}
T.~M. Svitkina, A.~B. Verkhovsky, K.~M. McQuade, and G.~G. Borisy, ``Analysis
  of the actin{\textendash}myosin {II} system in fish epidermal keratocytes:
  Mechanism of cell body translocation,'' {\em The Journal of Cell Biology},
  vol.~139, pp.~397--415, oct 1997.

\bibitem{ROSING1972275}
J.~Rosing and E.~Slater, ``The value of ${\Delta} {G}^o$ for the hydrolysis of
  {ATP},'' {\em Biochimica et Biophysica Acta (BBA) - Bioenergetics}, vol.~267,
  no.~2, pp.~275 -- 290, 1972.

\bibitem{PhyBioCell2012}
R.~Phillips, J.~Kondev, J.~Theriot, and H.~Garcia, {\em Physical Biology of the
  Cell}.
\newblock Garland Science, nov 2012.

\bibitem{cellbiobook}
R.~Milo and R.~Phillips, {\em Cell Biology by the Numbers}.
\newblock Garland Science, 2015.

\bibitem{dill2012molecular}
K.~Dill and S.~Bromberg, {\em Molecular driving forces: statistical
  thermodynamics in biology, chemistry, physics, and nanoscience}.
\newblock Garland Science, 2012.

\bibitem{Shelby_2003}
J.~P. Shelby, J.~White, K.~Ganesan, P.~K. Rathod, and D.~T. Chiu, ``A
  microfluidic model for single-cell capillary obstruction by plasmodium
  falciparum-infected erythrocytes,'' {\em Proceedings of the National Academy
  of Sciences}, vol.~100, pp.~14618--14622, nov 2003.

\bibitem{Plodinec_2012}
M.~Plodinec, M.~Loparic, C.~A. Monnier, E.~C. Obermann, R.~Zanetti-Dallenbach,
  P.~Oertle, J.~T. Hyotyla, U.~Aebi, M.~Bentires-Alj, R.~Y.~H. Lim, and C.-A.
  Schoenenberger, ``The nanomechanical signature of breast cancer,'' {\em
  Nature Nanotechnology}, vol.~7, pp.~757--765, oct 2012.

\bibitem{Pachenari_2014}
M.~Pachenari, S.~Seyedpour, M.~Janmaleki, S.~B. Shayan, S.~Taranejoo, and
  H.~Hosseinkhani, ``Mechanical properties of cancer cytoskeleton depend on
  actin filaments to microtubules content: Investigating different grades of
  colon cancer cell lines,'' {\em Journal of Biomechanics}, vol.~47,
  pp.~373--379, jan 2014.

\bibitem{Ketene_2012}
A.~N. Ketene, E.~M. Schmelz, P.~C. Roberts, and M.~Agah, ``The effects of
  cancer progression on the viscoelasticity of ovarian cell cytoskeleton
  structures,'' {\em Nanomedicine: Nanotechnology, Biology and Medicine},
  vol.~8, pp.~93--102, jan 2012.

\bibitem{MathematicaProgram}
I.~Wolfram~Research, {\em Mathematica}.
\newblock Wolfram Research, Inc., 2015.

\bibitem{Wolgemuth2011}
C.~W. Wolgemuth and P.~Lee, ``Crawling cells can close wounds without purse
  strings or signaling,'' {\em Biophysical Journal}, vol.~100, pp.~440a--441a,
  feb 2011.

\bibitem{Doxzen2013}
K.~Doxzen, S.~R.~K. Vedula, M.~C. Leong, H.~Hirata, N.~S. Gov, A.~J. Kabla,
  B.~Ladoux, and C.~T. Lim, ``{Guidance of collective cell migration by
  substrate geometry},'' {\em Integrative Biology}, vol.~5, no.~8, p.~1026,
  2013.

\bibitem{Pathak2012}
A.~Pathak, C.~S. Chen, A.~G. Evans, and R.~M. McMeeking, ``{Structural
  Mechanics Based Model for the Force-Bearing Elements Within the Cytoskeleton
  of a Cell Adhered on a Bed of Posts},'' {\em Journal of Applied Mechanics},
  vol.~79, no.~6, p.~061020, 2012.

\bibitem{Wang1994}
N.~Wang and D.~Ingber, ``Control of cytoskeletal mechanics by extracellular
  matrix, cell shape, and mechanical tension,'' {\em Biophysical Journal},
  vol.~66, pp.~2181--2189, jun 1994.

\bibitem{Wang_2001}
N.~Wang, I.~M. Tolic-Norrelykke, J.~Chen, S.~M. Mijailovich, J.~P. Butler,
  J.~J. Fredberg, and D.~Stamenovic, ``Cell prestress. i. stiffness and
  prestress are closely associated in adherent contractile cells,'' {\em {AJP}:
  Cell Physiology}, vol.~282, pp.~C606--C616, oct 2001.

\bibitem{Stamenovic_2002}
D.~Stamenovic and D.~E. Ingber, ``Models of cytoskeletal mechanics of adherent
  cells,'' {\em Biomechanics and Modeling in Mechanobiology}, vol.~1,
  pp.~95--108, jun 2002.

\bibitem{Han16689}
Y.~Han, S.~C. Cowin, M.~B. Schaffler, and S.~Weinbaum, ``Mechanotransduction
  and strain amplification in osteocyte cell processes,'' {\em Proceedings of
  the National Academy of Sciences}, vol.~101, no.~47, pp.~16689--16694, 2004.

\bibitem{Fungbook}
Y.-C. Fung, {\em Biomechanics}.
\newblock Springer New York, 1993.

\bibitem{Holzapfel2000art}
G.~A. Holzapfel, T.~C. Gasser, and R.~W. Ogden, ``A new constitutive framework
  for arterial wall mechanics and a comparative study of material models,''
  {\em Journal of elasticity and the physical science of solids}, vol.~61,
  pp.~1--48, Jul 2000.

\bibitem{Nappi_2015}
F.~Nappi, A.~R. Carotenuto, D.~D. Vito, C.~Spadaccio, C.~Acar, and M.~Fraldi,
  ``Stress-shielding, growth and remodeling of pulmonary artery reinforced with
  copolymer scaffold and transposed into aortic position,'' {\em Biomechanics
  and Modeling in Mechanobiology}, vol.~15, pp.~1141--1157, nov 2015.

\bibitem{Brangwynne_2006}
C.~P. Brangwynne, F.~C. MacKintosh, S.~Kumar, N.~A. Geisse, J.~Talbot,
  L.~Mahadevan, K.~K. Parker, D.~E. Ingber, and D.~A. Weitz, ``Microtubules can
  bear enhanced compressive loads in living cells because of lateral
  reinforcement,'' {\em J Cell Biol}, vol.~173, pp.~733--741, jun 2006.

\bibitem{Li_2008}
T.~Li, ``A mechanics model of microtubule buckling in living cells,'' {\em
  Journal of Biomechanics}, vol.~41, no.~8, pp.~1722--1729, 2008.

\bibitem{Soheilypour_2015}
M.~Soheilypour, M.~Peyro, S.~J. Peter, and M.~R. Mofrad, ``Buckling behavior of
  individual and bundled microtubules,'' {\em Biophysical Journal}, vol.~108,
  pp.~1718--1726, apr 2015.

\bibitem{plosone}
S.~Qin, V.~Ricotta, M.~Simon, R.~A.~F. Clark, and M.~H. Rafailovich,
  ``Continual cell deformation induced via attachment to oriented fibers
  enhances fibroblast cell migration,'' {\em PLOS ONE}, vol.~10, pp.~1--16, 03
  2015.

\bibitem{KAUNAS2009}
R.~Kaunas and H.-J. Hsu, ``A kinematic model of stretch-induced stress fiber
  turnover and reorientation,'' {\em Journal of Theoretical Biology}, vol.~257,
  no.~2, pp.~320 -- 330, 2009.

\bibitem{Qian_2013}
J.~Qian, H.~Liu, Y.~Lin, W.~Chen, and H.~Gao, ``A mechanochemical model of cell
  reorientation on substrates under cyclic stretch,'' {\em {PLoS} {ONE}},
  vol.~8, p.~e65864, jun 2013.

\bibitem{ANSYS}
{Ansys 15.0 User's Manual}, {\em ANSYS Mechanical User's Guide}.
\newblock ANSYS, Inc, release 15.0~ed., November 2013.

\bibitem{prsa2018}
S.~Palumbo, L.~Deseri, D.~Owen, and M.~Fraldi, ``Disarrangements and
  instabilities in augmented 1d hyperelasticity,'' {\em Proceedings of the
  Royal Society A: Mathematical, Physical and Engineering Sciences}, 2018.
\newblock In press.

\bibitem{FRALDI20181}
M.~Fraldi and A.~R. Carotenuto, ``Cells competition in tumor growth
  poroelasticity,'' {\em Journal of the Mechanics and Physics of Solids},
  vol.~112, pp.~345 -- 367, 2018.

\bibitem{CAROTENUTO20181}
A.~Carotenuto, A.~Cutolo, A.~Petrillo, R.~Fusco, C.~Arra, M.~Sansone,
  D.~Larobina, L.~Cardoso, and M.~Fraldi, ``Growth and in vivo stresses traced
  through tumor mechanics enriched with predator-prey cells dynamics,'' {\em
  Journal of the Mechanical Behavior of Biomedical Materials}, vol.~86, pp.~55
  -- 70, 2018.

\end{thebibliography}

\end{document}